\documentclass[bibyear]{aa}
\usepackage{graphicx}
\usepackage{txfonts}
\usepackage{lscape}

\newenvironment{itemlist}{
\begin{list}{}{\setlength{\leftmargin}{10mm}\setlength{\parsep}{0mm}
\setlength{\itemsep}{3mm}} }{\end{list}}
\usepackage{color}

\begin{document}

   \title{Far-infrared photometric observations of the outer planets and satellites with
          Herschel\thanks{{\it Herschel} is an ESA space observatory with science
          instruments provided by European-led Principal Investigator consortia and
          with important participation from NASA.}-PACS}

   \author{
          T. G. M\"{u}ller \inst{1},
          Z. Balog \inst{2},
          M. Nielbock \inst{2},
          R. Moreno \inst{3},
          U. Klaas \inst{2},
          A. Mo\'or \inst{4},
          H. Linz \inst{2},
          H. Feuchtgruber \inst{1}
          }

   \institute{   {Max-Planck-Institut f\"{u}r extraterrestrische Physik,
                 Postfach 1312, Giessenbachstra{\ss}e,
                 85741 Garching, Germany
                 }
          \and
                 {Max-Planck-Institut f\"{u}r Astronomie,
                 K\"onigstuhl 17, 69117 Heidelberg, Germany
                 }
          \and
                 {Observatoire de Paris, Laboratoire d'Etudes Spatiales
                 et d'Instrumentation en Astrophysique (LESIA),
                 5 Place Jules Janssen, 92195 Meudon Cedex, France
                 }
          \and
                 {Konkoly Observatory, Research Center for Astronomy and
                 Earth Sciences, Hungarian Academy of Sciences;
                 Konkoly Thege 15-17, H-1121 Budapest, Hungary
                 }
          }

   \date{Received ; accepted }

\abstract{We present all Herschel PACS photometer observations of \object{Mars},
\object{Saturn}, \object{Uranus}, \object{Neptune}, \object{Callisto},
\object{Ganymede}, and \object{Titan}. All measurements were carefully inspected for
quality problems, were reduced in a (semi-)standard way, and were calibrated. The derived flux densities
are tied to the standard PACS photometer response calibration, which is based on repeated
measurements of five fiducial stars. The overall absolute flux uncertainty is dominated by
the estimated 5\% model uncertainty of the stellar models in the PACS wavelength range between
60 and 210\,$\mu$m. A comparison with the corresponding planet and satellite models shows
excellent agreement for Uranus, Neptune, and Titan, well within the specified 5\%. Callisto
is brighter than our model predictions by about 4-8\%, Ganymede by about 14-21\%.
We discuss possible reasons for the model offsets.
The measurements of these very bright point-like sources, together with observations of stars and asteroids,
show the high reliability of the PACS photometer observations and the linear behavior of the PACS bolometer
source fluxes over more than four orders of magnitude (from mJy levels up to more than 1000\,Jy).
Our results show the great potential of using the observed solar system targets for cross-calibration purposes
with other ground-based, airborne, and space-based instruments and projects. At the same time, the PACS results will lead
to improved model solutions for future calibration applications.}


\authorrunning{M\"uller et al.}
\titlerunning{Far-infrared photometric observations of the outer planets and satellites with Herschel-PACS}
\maketitle


\section{Introduction}
\label{intro}

The planets of the solar system and their satellites beyond Earth have
surface and brightness temperatures of a few hundred Kelvin and are thereby
bright infrared emitters. With radiometers on space probes, such
as the
infrared radiometer for Mariner (Chase \cite{chase69}), IRIS on
Voyager (Hanel et al.\ \cite{hanel80}), and the Photopolarimeter-Radiometer (PPR)
on Galileo (Russell et al.\ \cite{russell92}), accurate information on
temperatures, thermal properties, albedo, energy balance, and infrared
emission spectra have been collected, making the outer planets and their
satellites suitable bright calibration standards. 
Radio occultation data acquired with Voyager have been used to probe the
vertical structure of the planetary atmospheres (Lindal \cite{lindal92}).
The inner planets are not accessible to cryogenic space telescopes, since they
are inside the solar constraint for these facilities.

In particular, Uranus and Neptune have been established as excellent flux 
standards for contemporary far-infrared space observatories, providing well-adapted flux 
levels to the dynamic range of their instrument detectors. Mars, 
Jupiter, and Saturn are too bright for sensitive photometers, but are 
observable with far-infrared instruments
with high spectral resolution. The 
SPIRE (Griffin et al.\ \cite{griffin10}) and PACS (Poglitsch et al.\
\cite{poglitsch10}) instruments onboard the Herschel Space Observatory
(Pilbratt et al.\ \cite{pilbratt10}) both used Uranus and Neptune, either as prime
calibrators in the case of SPIRE (Swinyard et al.\ \cite{swinyard10}), or as 
complementary calibrators to stellar (Dehaes et al.\ \cite{dehaes11}) and
asteroid (M\"uller et al.\ \cite{mueller14}) prime calibrators in the case
of PACS. For the latter flux calibration scheme, this approach covered a
wide flux range to address the non-linear response of the detectors and allowed
consistency checks in the case of overlapping fluxes of different calibrator
types and reference models. In addition, bright point sources allow obtaining measurements with
very high signal-to-noise ratio (S/N) with which different
instrument-specific aspects can be characterized, such as point-spread functions, cross-talk, and ghosts. They also serve to determine optimal observing strategies. For the PACS spectrometer, the bright
solar system targets helped to establish the in-flight relative spectral response
and wavelength calibration. 

In preparation of and during the Herschel mission, a number of workshops took
place selecting and reviewing the best-suited planet, satellite, 
asteroid, and stellar flux models, which were applied in support of the absolute
photometric calibration.

Here, we provide precise far-infrared photometry of the outer planets Uranus
and Neptune and for the planet satellites Callisto, Ganymede, and Titan
based on an independent calibration with stellar standards. We perform a
thorough comparison with best-suited models, from which feedback 
and constraints for future improvements is provided.

Our photometry is also a substantial input for cross-calibration aspects, in
particular to connect Herschel-PACS calibration with calibration schemes on
ground-based (sub-millimeter and millimeter facilities such as ALMA, APEX, or
IRAM/NOEMA), with airborne facilities (SOFIA), or other space telescope
facilities (ISO, Akari, Spitzer, and Planck).

In Sect.~\ref{sec:obs} we present the available PACS photometer chop-nod and
scan-map observations from the Herschel calibration and science programs.
Section~\ref{sec:drc} explains the data reduction, calibration, and flux
extraction procedures. In Sect.~\ref{sec:comparison} we present the observational
results and show comparisons with model predictions, in Sect.~\ref{sec:results} we discuss
the results, and in Sect.~\ref{sec:conclusions} we summarize our findings and
give a brief outlook. The observation details, instrument configurations,
quality aspects, and the Herschel-centric observing geometries are
provided in the appendix for completeness.

\section{Observations}
\label{sec:obs}

All measurements resulting in a derived flux presented here are taken from the
coordinated PACS calibration plan that was executed over the full mission.
Motivation for and classification of
individual measurements with respect to\ the overall calibration strategy are described in
the PACS Calibration document\footnote{PCD; PACS-MA-GS-001, Issue 1.10, Nov.\ 2014.},
the respective PACS Performance Verification Phase Plan\footnote{PICC-MA-PL-001, Issue 2.0, May 2014.},
and the PACS Routine (Science) Phase Calibration Plan\footnote{PICC-MA-PL-002, Issue 4.01, May 2014.},
all available from the Herschel Explanatory Library Listings (HELL\footnote{\tt http://www.cosmos.esa.int/\-web/\-herschel.}).

The PACS photometer exploited two filled silicon bolometer arrays with 32$\times$64 pixels (blue camera) and
16$\times$32 pixels (red camera)
to perform imaging photometry in the 60 - 210\,$\mu$m wavelength range. Observations were
taken simultaneously in two bands, 60 - 85\,$\mu$m (blue band with $\lambda_{ref}$ = 70.0\,$\mu$m)
or 85 - 130\,$\mu$m (green band with $\lambda_{ref}$ = 100.0\,$\mu$m) and
130 - 210\,$\mu$m (red band with $\lambda_{ref}$ = 160.0\,$\mu$m), over
a field of view of 1.75$^{\prime}$ $\times$ 3.5$^{\prime}$, with full beam sampling
in each band. Technical details are given in Poglitsch et al.\ (\cite{poglitsch10}), the
description of the chop-nod and scan-map observing techniques are presented in Nielbock
et al.\ (\cite{nielbock13}) and Balog et al.\ (\cite{balog14}), respectively, in the
PACS Observer's Manual\footnote{\tt http://herschel.esac.esa.int/Docs/PACS/html/\-pacs\_om.html}
(Altieri et al.\ \cite{pacsom}), and references therein.

\subsection{PACS chop-nod observations}

The originally recommended PACS photometer observing
mode for point and compact sources was the chop-nod point-source photometry mode.
This mode used the PACS chopper to move the source by about 50$^{\prime \prime}$, corresponding
to the size of about 1 blue/green bolometer matrix (16 pixels) or the size of about half a red matrix (8 pixels),
with a chopper frequency of 1.25\,Hz. The nodding is performed by a satellite movement of the same amplitude, but
perpendicular to the chopping direction. On each nod-position the chopper executed 3$\times$25 chopper cycles.
The three sets of chopper patterns are either on the same array positions (no dithering) or on three different array
positions (dither option). Our bright-source observations here were all made with the dither option, where the chopper
pattern was displaced in $\pm$Y-direction (along the chopper direction) by 8.5$^{\prime \prime}$
(2 2/3 blue pixels or 1 1/3 red pixels). Each chopper plateau lasted for 0.4\,s (16 readouts on-board),
producing four frames per plateau in the down-link. The full 3$\times$25 chopper cycles per nod-position
were completed in less than one minute. The pattern was repeated on the second nod-position. When repetition factors exceeded 1 (usually only for fainter targets),
the nod-cycles were repeated in the following way (example for four repetitions):
nodA-nodB-nodB-nodA-nodA-nodB-nodB-nodA to minimise satellite slew times.
Our chop-nod observations were taken either in low or high gain. More details about PACS photometer
observations taken in chop-nod mode can be found in Nielbock et al.\ (\cite{nielbock13}).
Our PACS chop-nod observations are listed in Table~\ref{tbl:pacs_chopnod} in the appendix.

\subsection{PACS scan-map observations}

Although originally not designed for point-source observations, the scan-map technique
replaced the point-source chop-nod mode. After the science demonstration phase (SDP) it
was recommended to use a so-called mini scan-map mode for observations of point and
compact sources. The mini scan-map mode had a higher sensitivity and
allowed a better characterization of the close vicinity of the target and
larger scale structures in the background.
Most of the bright-source measurements were executed in the mini scan-map implementation
as recommended in the official
release note\footnote{\tt http://herschel.esac.esa.int/twiki/pub/Public/\-PacsCalibrationWeb/\-PhotMiniScan\_ReleaseNote\_20101112.pdf}.
The satellite scans were usually made with the nominal 20$^{\prime \prime}$/s speed (see exceptions in Sect.~\ref{sec:quality}
and in Table~\ref{tbl:pacs_scanmap1} in the appendix) in array coordinates of 70$^{\circ}$ and 110$^{\circ}$ (along the diagonal of the bolometer arrays).
Only a few early measurements were made under different angles of 63$^{\circ}$ and 117$^{\circ}$ or 45$^{\circ}$ and 135$^{\circ}$. 
The scan-map observations have different scan-leg lengths of 2.0$^{\prime}$ to 7.0$^{\prime}$,
but usually a total of 10 legs and a separation of 4$^{\prime \prime}$ between the scan legs.
The repetition factor for the mini scan-maps was 1, with the exception of two sets of Neptune measurements
where the scan-maps were repeated four times. Some of the scan-map observations were taken in low gain to avoid saturation.
Mars was taken in high gain (despite its very high flux) with the goal to
characterize the point-spread function (PSF) wings. Here the measurements
suffer from saturation in the PSF core and a standard flux determination through aperture photometry
is not possible. Another two sets of measurements are related to Saturn and Uranus (see Table~\ref{tbl:pacs_scanmap2}), but the 
planets were located at the edge or even outside the observed field. These measurements are listed, but a flux determination
for the planets was not possible. A full description of the scan-map mode and its performance can be found in 
Balog et al.\ (\cite{balog14}).
The observational details of our PACS scan-map observations are listed in Tables~\ref{tbl:pacs_scanmap1} and
\ref{tbl:pacs_scanmap2}.

\subsection{Observation quality information}
\label{sec:quality}

Most of the PACS photometer science and calibration measurements
are of very high quality. Only a few measurements suffered from
suboptimal instrument settings, instrument or satellite events, or
space environment influences. All problematic and quality-related issues
are collected in various reports and documents, available from the HELL.

Our sample has two sets of measurements with pointing-critical
solar aspect angles (SAA) of the spacecraft:
(1) a set of four scan-map observations of Neptune in OD 759 took
place at almost -20$^{\circ}$ SAA; (2) the complete set of four Titan
measurements was executed on OD 1138 at -18.3$^{\circ}$ SAA.
A degradation of the pointing performance has been measured for these
warm attitudes (S\'anchez-Portal et al.\ \cite{sanchezportal14}),
but a correction is not possible at this stage. The astrometry of these
Neptune and Titan measurements is therefore less reliable, but there
was no influence on the aperture photometry, which was performed directly at the location
of the bright source in the final maps.

The PACS photometer calibration is mainly based on observations taken in the nominal satellite scan speed of
20$^{\prime \prime}$/s. However, some of the planet observations were
executed either with 10$^{\prime \prime}$/s or 60$^{\prime \prime}$/s
scan speed (see Table~\ref{tbl:pacs_scanmap1}).
Point-source fluxes in different
sky fields and based on different satellite scan speeds were analyzed, but no obvious problems  were found (see the report on the PACS map-making
tools: analysis and benchmarking\footnote{\tt http://herschel.esac.esa.int/twiki/pub/Public/\-PacsCalibrationWeb/\-pacs\_mapmaking\_report\_ex\_sum\_v3.pdf}).
The detector response calibration is not affected by the satellite scan speed,
but the reliability of the aperture photometry might be slightly
reduced for high scan speeds of 60$^{\prime \prime}$/s
by small differences of a few percent in the encircled energy fraction
(EEF) within a given aperture (see also
PICC-ME-TN-033\footnote{\tt http://herschel.esac.esa.int/twiki/pub/Public/\-PacsCalibrationWeb/bolopsf\_21.pdf},
Vers.\ 2.1 or later). In these cases it is recommended to use relatively
large apertures to include the slightly wider PSF. This was done here for
our bright sources. We note that specific EEF tables for 60$^{\prime \prime}$/s
scan speed are in preparation.

Most of our observations have standard instrument settings for
chop-nod or scan-map modes; these measurements can easily
be found in the HSA, and the pipeline provides reliable products.
Only the two observations of Callisto and Ganymede were taken in non-standard
settings and the HSA postcards are missing, but the reliability of
the corresponding photometry is not affected (see discussion in
Sect.~\ref{sec:results}).

Four Neptune measurements from OD 1097 on May 15, 2012 (OBSIDs 1342245787,
1342245788, 1342245789, and 1342245790) suffered from a failure of the blue
SPU\footnote{Signal Processing Unit, part of the PACS warm electronics.}
 ; they are labeled "FAILED" in the HSA. The measurements in the 
red, long-wavelength channel are very likely not affected by this event,
but were not available at the time of processing and therefore were
not included in our analysis.

After OD 1375 (February 17, 2013) half of the red PACS photometer array was lost
(indicated as "red matrix saturated"), but point-source photometry was still possible.
Four Neptune measurements from OD 1444 (April 26, 2013; OBSIDs 1342270939, 1342270940,
1342270941, and 1342270942) were affected, but the data are clean otherwise, and aperture
photometry was possible in both channels for all four maps.

\section{Data reduction, calibration, and photometry}
\label{sec:drc}

\subsection{Processing of chop-nod and scan-map observations}
\label{sec:reduction}

The data reduction and calibration of chop-nod and scan-map data
is described in Nielbock et al.\ (\cite{nielbock13}) and Balog et
al.\ (\cite{balog14}), respectively. Specific aspects in the analysis
of moving solar system targets are addressed in Kiss et al.\
(\cite{kiss14}). To reduce our bright targets,
we adjusted a few settings and used very recent software developments
for PACS photometer observations. The new corrections and reduction
steps are meanwhile part of the standard product generation (SPG)
pipelines version 13.0 and higher:

\begin{itemlist}
\item[$\bullet$] Gyro correction: the latest satellite pointing products
                 include corrections for high-frequency small pointing
                 jitter on the basis of the satellite-internal gyros.
\item[$\bullet$] New calibration file for the focal plane geometry: correcting
                 for very small distortion effects that have not been handled
                 before.
\item[$\bullet$] Flux correction function for chop-nod observations (as given
                 in Nielbock et al.\ \cite{nielbock13}).
\item[] \underline{Only for scan-map observations:}
\item[$\bullet$] Precise timing of the sequential readout of the individual detector columns:
                 the photometer detectors are read out sequentially column by column during
                 a 40\,ms cycle. Combined with a satellite scan pattern of 10, 20, or 60$^{\prime \prime}$/s,
                 this correction improves the sharpness of the PSF by assigning more
                 accurate pointing information to each individual pixel.
\item[$\bullet$] Larger size of the mask (60$^{\prime \prime}$ instead
                 of 25$^{\prime \prime}$ given in Balog et al.\ \cite{balog14})
                 to account for the much more extended PSFs for the bright
                 sources: This mask is used in the context of the high-pass
                 filtering of the data to avoid flux losses for point sources.
\item[$\bullet$] High-pass filtering with a filter width of
                 15, 20, and 35 readouts in blue, green, and red band, respectively.
\item[$\bullet$] Frame selection based on scan speed (must be within $\pm$10\% of the
                 nominal speed).
\item[$\bullet$] Final projection of all data with {\tt photProject()}, using the
                 default pixel fraction (pixfrac = 1.0) and reduced map pixel sizes
                 of 1.1$^{\prime \prime}$, 1.4$^{\prime \prime}$, and 2.1$^{\prime \prime}$
                 in the blue, green, and red channel, respectively.
\end{itemlist}

For each target, each band, and each scan- and cross-scan as well as for each
chop-nod data set, we produced calibrated standard data products (maps) in the
object-centric reference system.

\subsection{Flux extraction and uncertainty estimation}
\label{sec:phot}

We performed standard aperture photometry (source flux and 1-$\sigma$ uncertainty) on each of the
final maps. Standard aperture radii of 12$^{\prime \prime}$, 12$^{\prime \prime}$,
and 22$^{\prime \prime}$ were used, requiring aperture correction factors of 0.802, 0.776,
and 0.817 at 70, 100, and 160\,$\mu$m, respectively.

\subsubsection{Chop-nod measurements}

The chop-nod observing technique eliminates the background automatically.
The photometric uncertainty was estimated from the fluctuations in a 
given sky annulus (see details in Nielbock et al.\ \cite{nielbock13}). Correlated noise is
corrected by an empirical function to obtain a conservative upper limit for the measurement
uncertainties. The relatively large uncertainties can be attributed to the chopping amplitude
and the field of view in chop-nod mode observations; both are too small and do not allow
determining clean background fields around our very bright sources. Typical flux uncertainties
(without the given 5\% absolute flux uncertainty from the detector response calibration
through prime standard stars) are well below 1\% in blue and green, and well below 2\% in red.

\subsubsection{Scan-map observations}

The sources are extremely bright, and
a combination of scan- and cross-scan measurements is not needed. We therefore present
the extracted fluxes for each observation (OBSID) separately.
To estimate the photometric uncertainty in scan-map measurements, we placed apertures with radii of 10$^{\prime \prime}$
in a 7 by 7 grid on the map around the source and measured the underlying fluxes. Then we used $\sigma$-clipping
to remove the apertures that were contaminated by the source flux (about half of the apertures
were eliminated). The final 1-$\sigma$ uncertainty is then the r.m.s.\ of fluxes from the remaining
blank apertures. Typical flux uncertainties (without the given 5\% absolute flux uncertainty
from the detector response calibration through selected stars) are well below 1\% in all three bands.

\subsection{Non-linearity correction}

Bolometers are thermal detectors where a thermistor converts radiation
(or heat) into an electrical signal. The impedance of the thermistor
strongly depends on its temperature, and the relation between incoming
flux and output voltage is non-linear. During the ground tests of the
PACS instrument, this non-linear behavior of the bolometers was extensively
characterized for a wide range of incoming fluxes. The non-linearity effects of each
individual pixel were fit by simple functions over the relevant range
of fluxes (telescope and sky combined) around the in-flight operating
point of the bolometers (Billot \cite{billot11}). 
These functions are used in the general processing steps of the PACS
bolometer to linearize the fluxes: the measured flux in a given
pixel and in a given band is multiplied with the corresponding
non-linearity correction factor. The
correction step is automatically applied by the photometer pipeline
in the level 1 data product generation, together with saturation,
flat-field, offset, and response calibration. 

The non-linearity correction factors are close to 1.0 for almost
all photometer observations. This correction is relevant only for the
brightest regions in the sky and very bright point sources.
For bright point sources - like the calibration asteroids - this
correction is between 0 and approximately 6\%, depending on the
source, band, and pixel (see M\"uller et al.\ \cite{mueller14}).
For the planets Uranus and Neptune, the correction increases to
about 5-15\%, while for Callisto and Ganymede the non-linearity
factor is 1.17 in the blue band and about 1.05 in the red band.

\subsection{Color corrections}

For the calculation of the monochromatic flux densities at the PACS photometer
reference wavelengths of 70.0\,$\mu$m, 100.0\,$\mu$m, and 160.0\,$\mu$m, it is
still required to perform a color correction to account for the difference of a
constant energy spectrum $\nu$$\cdot$F$_{\nu}$ = $\lambda$$\cdot$F$_{\lambda}$ = const.
(assumption in the PACS photometer calibration) and the true spectral energy
distribution (SED) of the object (M\"uller et al.\ \cite{mueller11}, Balog et al.\ \cite{balog14}).
Our calculated color-correction factors are listed in Table~\ref{tbl:cc}. These correction
factors and their estimated uncertainties are based on the listed model spectra.

\begin{table*}
  \caption{Color-correction factors (cc) for our bright sources. Models are taken
           from {\tt ftp://ftp.sciops.esa.int/\-planets/\-originalData/}. The
           estimated maximal uncertainty for these corrections is 2\% for Titan
           and about 1\% for the rest of the targets.}
  \label{tbl:cc}
  \begin{tabular}{llllll}
  \noalign{\smallskip}\hline\hline\noalign{\smallskip}
  Source & cc$_{70.0\,\mu m}$ & cc$_{100.0\,\mu m}$ & cc$_{160.0\,\mu m}$ & unc.\ [\%] & model source \\
  \noalign{\smallskip}\hline\noalign{\smallskip}
  Mars     & 0.947 & 1.033 & 1.054 & $\approx$1 & mars\_esa\_2\_i.dat (Moreno) \\
  Uranus   & 0.984 & 0.995 & 1.018 & $\approx$1 & orton\_uranus\_esa5 (Orton) \\  
  Uranus   & 0.984 & 0.992 & 1.019 & $\approx$1 & ura\_esa2\_2\_i.dat (Moreno) \\
  Neptune  & 0.984 & 0.993 & 1.020 & $\approx$1 & nep\_esa5\_2\_i.dat (Moreno)\\
  Callisto & 0.998 & 1.016 & 1.055 & $\approx$1 & call\_esa2\_2\_i.dat (Moreno) \\
  Ganymede & 0.996 & 1.014 & 1.053 & $\approx$1 & gany\_esa2\_2\_i.dat (Moreno) \\
  Titan    & 0.985 & 1.002 & 1.028 & $\approx$2 & tit\_esa3\_2\_i.dat (Moreno) \\
  \noalign{\smallskip}
  fiducial stars & 1.016 & 1.033 & 1.074 & $\approx$1 & Balog et al.\ (\cite{balog14}) \\
  \noalign{\smallskip}\hline
  \end{tabular}
\end{table*}

In the following tables we present the measured and calibrated fluxes together with the
monochromatic flux densities after color-correction at the PACS reference wavelengths.
The flux uncertainties presented in the tables are measurement errors and do not
include the uncertainties in color correction (estimated to
be $\approx$1\% in most cases, and $\approx$2\% for Titan showing
many strong lines in the PACS range) nor the absolute flux calibration
errors of 5\% (see Balog et al.\ \cite{balog14}). These have to
be added quadratically when using the derived fluxes in an absolute
sense.

\section{Observational results and comparison with model predictions}
\label{sec:comparison}

\begin{figure}[h!tb]
\centering
  \rotatebox{180}{\resizebox{\hsize}{!}{\includegraphics{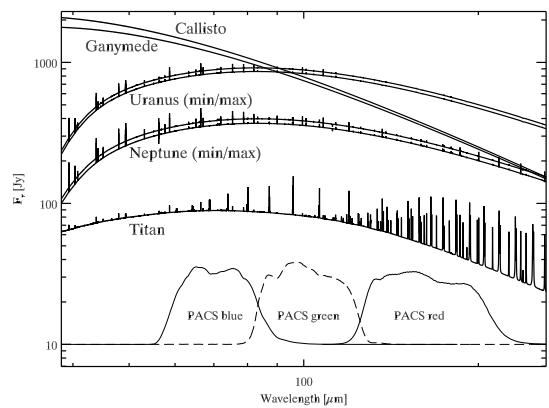}}}
  \caption{Absolute disk-integrated model flux density predictions for Callisto,
           Ganymede, and Titan in the Herschel-centric reference system and at
           the epoch of the corresponding PACS measurements.
           The minimum-maximum model predictions for Uranus ("ura\_esa2\_2\_i.dat")
           and Neptune refer to all available PACS measurements during the
           entire Herschel mission. The PACS band-passes are shown in arbitrary
           units. The blue, green, and red bands have peak transmissions
           of 51\%, 56\%, and 46\%, respectively.
\label{fig:sed_all}}
\end{figure}

The planet and satellite models were calculated from disk-averaged
brightness temperature spectra based on planetary atmosphere
models together with the Herschel-centric apparent solid angle
at the time of the observations. Various model versions were prepared
in support for the Herschel mission by model experts and provided to the
Herschel Calibration Steering Group. Specific model versions have then
been used by the three instrument teams for observation planning and specific
calibration purposes. Figure~\ref{fig:sed_all} shows the model predictions
for Uranus, Neptune, Callisto, Ganymede, and Titan either at the epoch of
the Herschel-PACS measurement or - for Uranus and Neptune - as
minimum-maximum prediction for all observing epochs.

\subsection{Mars}

The main goal of the Mars observations with the PACS photometer was to
characterize the PSF wings and to probe for ghosts and stray-light
features in the field of view (see D.\ Lutz \cite{lutz12},
PICC-ME-TN-033\footnote{\tt http://herschel.esac.esa.int/twiki/pub/Public/\-PacsCalibrationWeb/bolopsf\_21.pdf},
Vers.\ 2.1). The data were therefore taken in high-gain. The scan legs are much longer
and the leg separation much wider than in standard mini scan-map observations
in order to map out the PSF characteristics at large distances.
The central source parts (several arcsec wide) are flagged as completely
saturated in all final maps\footnote{Using a pre-Herschel delivery of a Mars model
"mars\_esa\_2\_i.dat",
combined with the true Herschel-centric diameter of Mars
gives model flux predictions of $\approx$44\,000 to 54\,000\,Jy at 70\,$\mu$m,
$\approx$25\,000 to 30\,000\,Jy at 100\,$\mu$m,
and $\approx$11\,000 to 14\,000\,Jy at 160\,$\mu$m
for the Herschel observing epochs.}.
For completeness reasons we present the
detailed observation information in Table~\ref{tbl:pacs_scanmap1} in the appendix.

\subsection{Uranus}
\label{sec:uranus}

\begin{figure}[h!tb]
  \rotatebox{90}{\resizebox{!}{\hsize}{\includegraphics{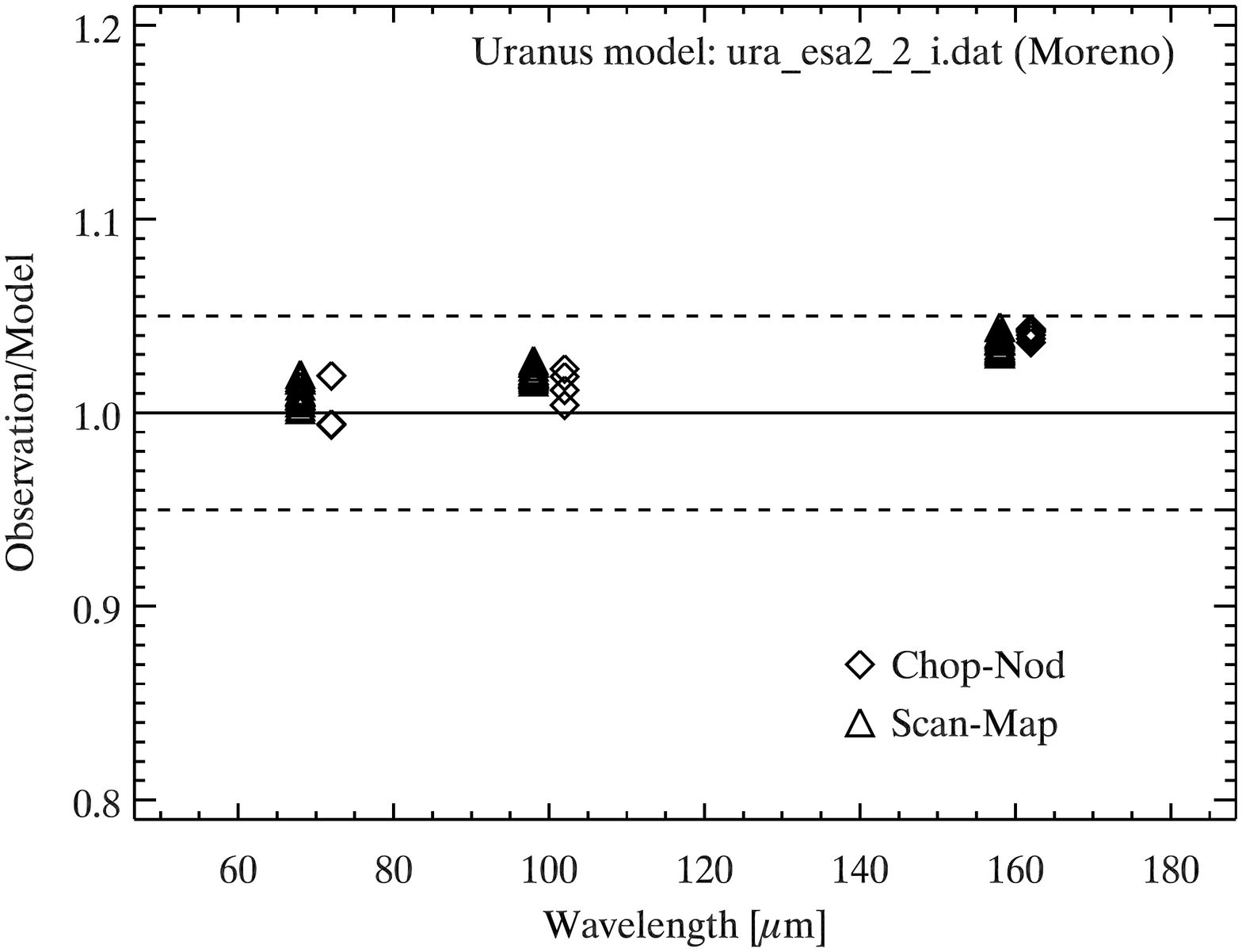}}}
  \rotatebox{90}{\resizebox{!}{\hsize}{\includegraphics{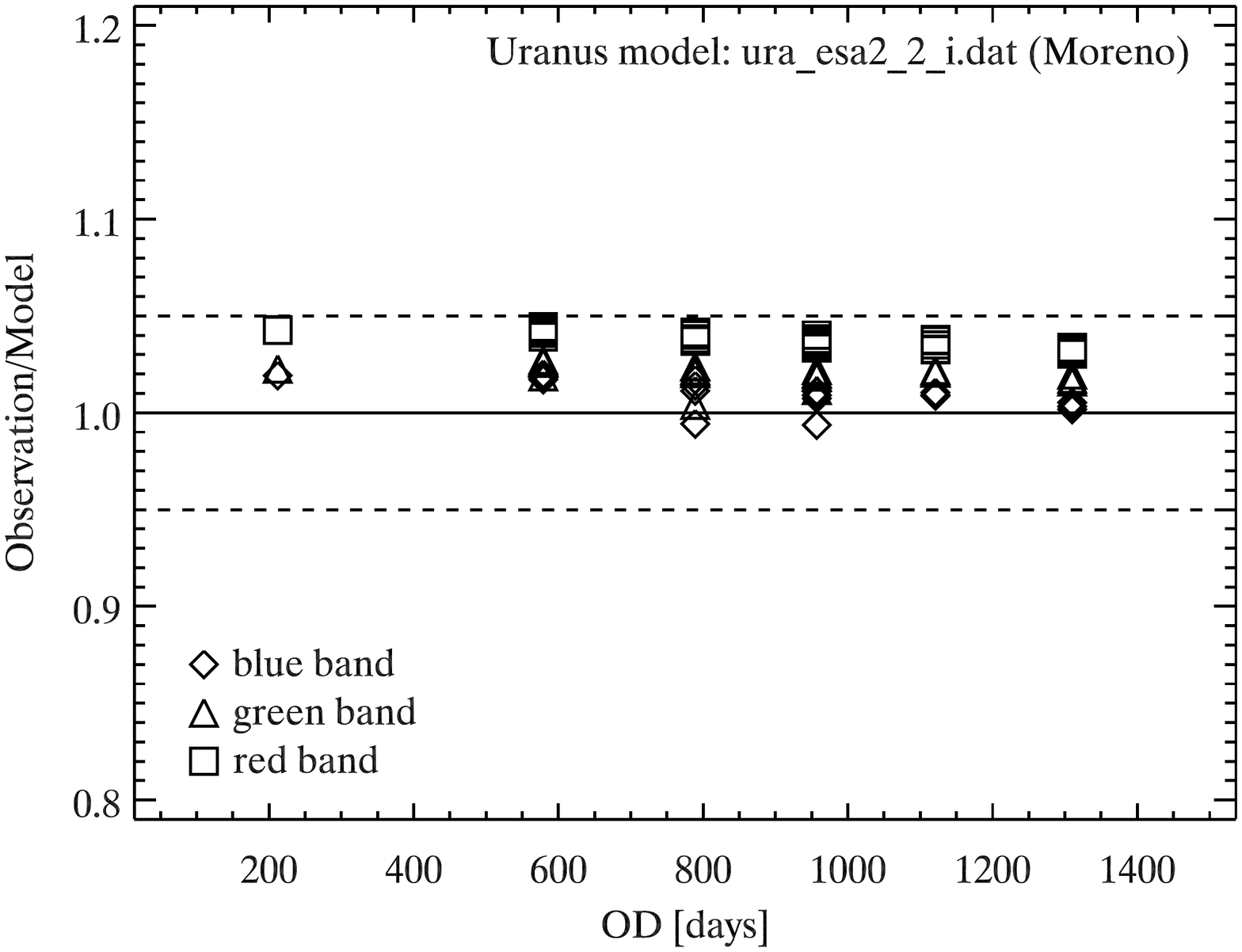}}}
  \caption{All PACS photometer observations of Uranus divided by the
           corresponding model predictions by R.\ Moreno. Top: as a function of wavelength
           (chop-nod and scan-map observations are shown with different symbols).
           Bottom: as a function of OD.
\label{fig:uranus1}}
\end{figure}

\begin{figure}[h!tb]
  \rotatebox{90}{\resizebox{!}{\hsize}{\includegraphics{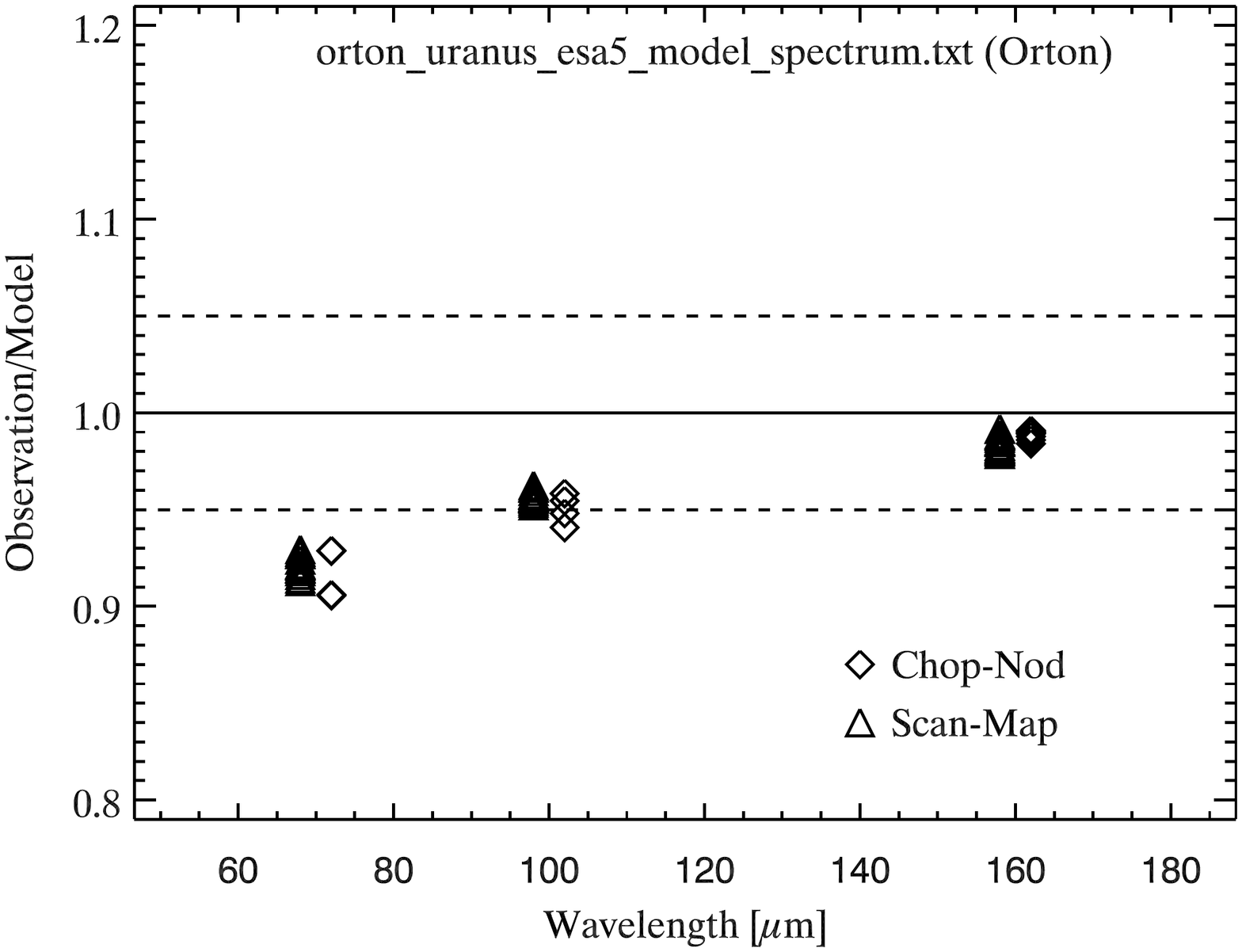}}}
  \rotatebox{90}{\resizebox{!}{\hsize}{\includegraphics{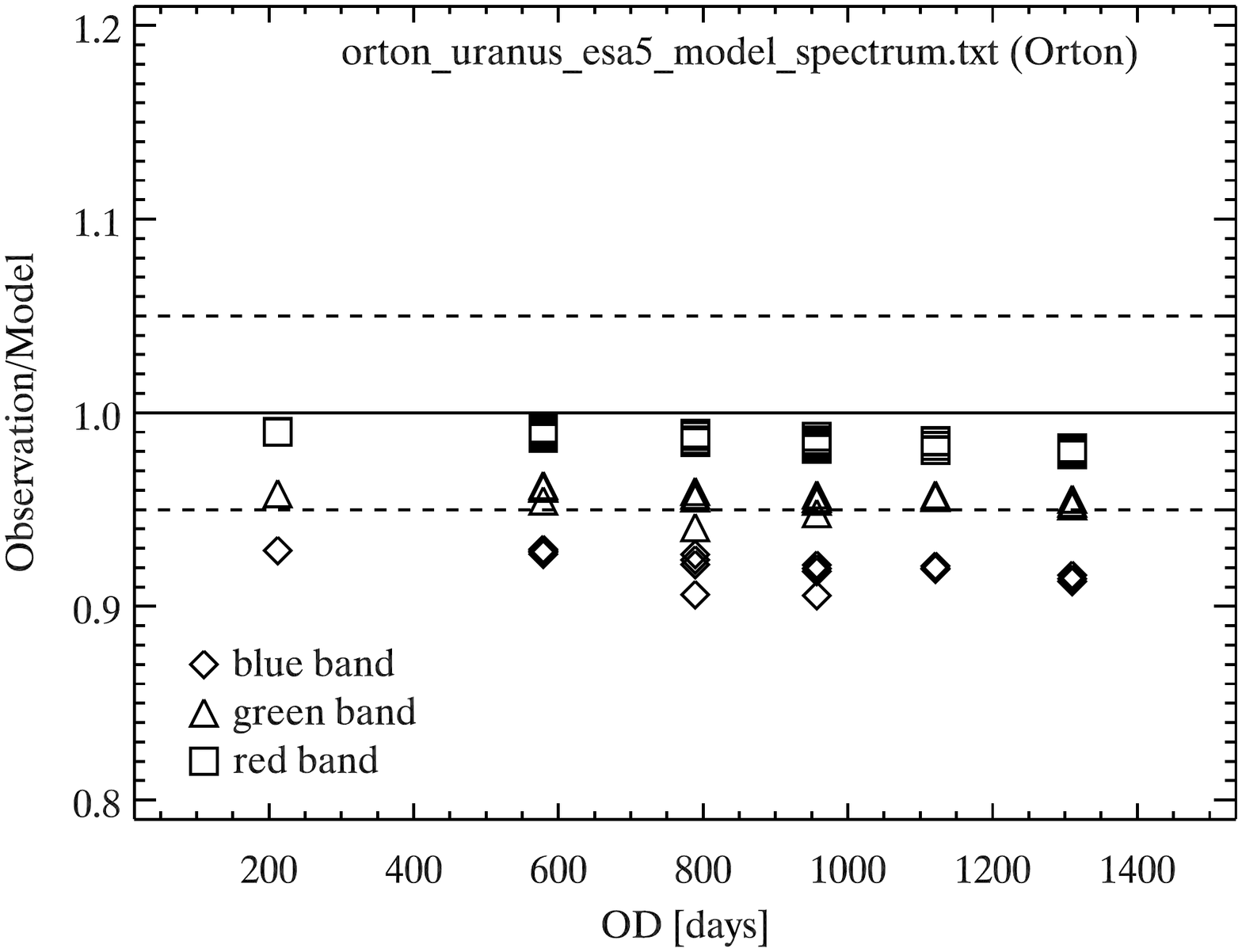}}}
  \caption{All PACS photometer observations of Uranus divided by the
           corresponding model predictions by G.\ Orton. Top: as a function of wavelength
           (chop-nod and scan-map observations are shown with different symbols).
           Bottom: as a function of OD.
\label{fig:uranus2}}
\end{figure}

Table~\ref{tbl:uranus} contains the reduced and calibrated fluxes of 
all Uranus observations obtained with the Herschel-PACS photometer. Each line
is connected to an output map that is related to a single OBSID based either on a
single scan direction or a single chop-nod observation. The observing geometry and relevant
instrument and satellite configuration parameters are given in Tables~\ref{tbl:pacs_chopnod}
and \ref{tbl:pacs_scanmap1}. The absolute calibrated flux densities (column "Flux") are
based on a flat energy spectrum ($\nu$F$_{\nu}$ = const.) with uncertainties
coming from the signal processing and aperture photometry of the final maps.
The color-corrected flux densities at the 
PACS reference wavelength are given in column $FD_{cc}$ , where color-correction factors
from Table~\ref{tbl:cc} were used. The model predictions are based on two different
models that are both relevant for different purposes in the context of PACS and
SPIRE photometer and spectrometer calibration: (i) the Uranus model
"orton\_uranus\_esa5\_model\_spectrum.txt", which was taken from the
HCalSG web pages and was provided by Glenn Orton, JPL/NASA (Orton et al.\ \cite{orton14});
(ii) the Uranus model  "ura\_esa2\_2\_i.dat", which was taken from the
HCalSG web pages at {\tt ftp://ftp.sciops.esa.int/\-planets/\-ori\-ginal\-Data/\-esa2/}
and was provided by R.\ Moreno in 2009 (see also Moreno et al.\ 2016, in preparation).
Both models of Uranus were used to check the PACS calibration. They are connected to
slightly different thermal structures: for Uranus "esa2" (available at the beginning
of the Herschel mission) the thermal profiles go back to Voyager radio-occultation
measurements (Pearl et al.\ \cite{pearl90}; Lindal et al.\ \cite{lindal92}),
while for Uranus "esa5" measurement constraints from mid- and far-IR
(Spitzer-IRS and Herschel/SPIRE; Orton et al.\ \cite{orton14}) were also considered. 
The radiative transfer modeling includes continuum opacities from collision-induced absorption of H$_2$, He, and CH$_4$.
"esa5" is more tuned toward the SPIRE wavelengths and was not tested so far
against PACS data. The brightness temperature differences
between the two models are below 5\% in the PACS wavelength range.
The model uncertainties are linked mainly to the input parameters and therefore
reflect the absolute calibration accuracy.
The model brightness temperatures were combined with Uranus' angular diameter
as seen from Herschel, taking the equatorial
and polar radii and the sub-observer latitude (i.e.,\ Ob-lat in JPL/Horizon)
into account\footnote{It is worth noting that the angular diameter values from the JPL-Horizons
system correspond to the equatorial diameters, which are about 0.7\% (Neptune) and
1.0\% (Uranus) larger than our calculated geometric diameters, and the JPL values would produce
too high model fluxes ($\approx$2\% too high for Uranus and $\approx$1.4\% too
high for Neptune).}.
The last column of Table~\ref{tbl:uranus}
contains the calculated ratio between $FD_{cc}$ and the corresponding model prediction
at the given reference wavelength.
The final observation-to-model ratios are shown in Figs.~\ref{fig:uranus1} (model by Moreno)
and \ref{fig:uranus2} (model by Orton). The two PACS observing modes - chop-nod and scan-map
techniques - are shown with different symbols and at slightly shifted wavelengths ($\pm$2\,$\mu$m)
for better visibility. The absolute uncertainties for both Uranus models were estimated to be 5\%,
similar to the stellar model uncertainties for the five fiducial stars.
The $\pm$5\% boundaries are shown as dashed lines in Figs.~\ref{fig:uranus1}, \ref{fig:uranus2},
\ref{fig:neptune}, and \ref{fig:callisto_ganymede_titan}.

{\small
\begin{longtab}
\begin{longtable}{rlcrrrrllll}
\caption[]{Observational results of the Herschel-PACS photometer observations of Uranus and comparison
           with two different models (see Text).} \\
\hline\noalign{\smallskip}
   &       &                   & PACS & Flux & Unc.\ & $\lambda_{ref}$ & FD$_{cc}$ & FD$_{moreno}$ & \multicolumn{2}{c}{Ratio FD$_{cc}$/FD$_{model}$} \\
OD & OBSID & observ.\ mid-time & Band & [Jy] & [Jy]  & [$\mu$m]        & [Jy]      & [Jy]         & Moreno & Orton \\
\noalign{\smallskip}\hline\noalign{\smallskip}
\multicolumn{11}{l}{chop-nod:} \\
\endfirsthead                       
\caption[]{\emph{continued}} \\                                                     
\hline\noalign{\smallskip}                                                                                  
   &       &                   & PACS & Flux & Unc.\ & $\lambda_{ref}$ & FD$_{cc}$ & FD$_{moreno}$ & \multicolumn{2}{c}{Ratio FD$_{cc}$/FD$_{model}$} \\
OD & OBSID & observ.\ mid-time & Band & [Jy] & [Jy]  & [$\mu$m]        & [Jy]      & [Jy]         & Moreno & Orton \\
\noalign{\smallskip}\hline\noalign{\smallskip}  \endhead                                                    
\hline                                                                                                      
\multicolumn{11}{c}{\emph{continued on next page}} \endfoot                                                 
\noalign{\smallskip}\hline \endlastfoot                                                                     
\label{tbl:uranus}
 212 & 1342188056                & 2455178.12069 & B & 864.812 &   0.486 &  70.0 & 878.874 & 862.299 & 1.02 & 0.93 \\       
 212 & 1342188056                & 2455178.12069 & R & 657.711 &   9.139 & 160.0 & 645.447 & 619.000 & 1.04 & 0.99 \\       
 212 & 1342188057                & 2455178.12340 & G & 873.153 &   1.815 & 100.0 & 880.195 & 860.737 & 1.02 & 0.96 \\       
 212 & 1342188057                & 2455178.12340 & R & 657.576 &   9.022 & 160.0 & 645.315 & 619.000 & 1.04 & 0.99 \\       
 579 & 1342211116                & 2455544.21598 & B & 870.285 &   1.572 &  70.0 & 884.436 & 867.964 & 1.02 & 0.93 \\       
 579 & 1342211116                & 2455544.21598 & R & 661.492 &   9.246 & 160.0 & 649.158 & 623.067 & 1.04 & 0.99 \\       
 579 & 1342211119                & 2455544.22774 & G & 875.409 &   1.798 & 100.0 & 882.469 & 866.343 & 1.02 & 0.95 \\       
 579 & 1342211119                & 2455544.22774 & R & 662.397 &   8.880 & 160.0 & 650.046 & 623.031 & 1.04 & 0.99 \\       
 789 & 1342223981                & 2455754.55343 & B & 866.861 &   1.521 &  70.0 & 880.956 & 886.077 & 0.99 & 0.91 \\       
 789 & 1342223981                & 2455754.55343 & R & 672.918 &   8.103 & 160.0 & 660.371 & 636.069 & 1.04 & 0.99 \\       
 789 & 1342223984                & 2455754.56491 & G & 880.840 &   1.638 & 100.0 & 887.944 & 884.473 & 1.00 & 0.94 \\       
 789 & 1342223984                & 2455754.56491 & R & 675.109 &   8.154 & 160.0 & 662.521 & 636.069 & 1.04 & 0.99 \\       
 957 & 1342235628                & 2455922.43656 & B & 837.225 &   1.405 &  70.0 & 850.838 & 856.257 & 0.99 & 0.91 \\       
 957 & 1342235628                & 2455922.43656 & R & 649.040 &   8.481 & 160.0 & 636.938 & 614.662 & 1.04 & 0.98 \\       
 957 & 1342235631                & 2455922.45609 & G & 857.702 &   1.480 & 100.0 & 864.619 & 854.657 & 1.01 & 0.95 \\       
 957 & 1342235631                & 2455922.45609 & R & 651.405 &   8.372 & 160.0 & 639.259 & 614.627 & 1.04 & 0.99 \\       
\noalign{\smallskip}
\multicolumn{11}{l}{scan-map:} \\
 579 & 1342211117                & 2455544.22060 & B & 871.064 &   1.869 &  70.0 & 885.228 & 867.964 & 1.02 & 0.93 \\       
 579 & 1342211118                & 2455544.22464 & B & 868.696 &   1.858 &  70.0 & 882.821 & 867.964 & 1.02 & 0.93 \\       
 579 & 1342211120                & 2455544.23128 & G & 882.701 &   1.504 & 100.0 & 889.820 & 866.343 & 1.03 & 0.96 \\       
 579 & 1342211121                & 2455544.23532 & G & 881.411 &   1.770 & 100.0 & 888.519 & 866.343 & 1.03 & 0.96 \\       
 579 & 1342211117                & 2455544.22060 & R & 659.878 &   1.674 & 160.0 & 647.574 & 623.067 & 1.04 & 0.99 \\       
 579 & 1342211118                & 2455544.22464 & R & 662.876 &   1.508 & 160.0 & 650.516 & 623.067 & 1.04 & 0.99 \\       
 579 & 1342211120                & 2455544.23128 & R & 660.905 &   1.646 & 160.0 & 648.582 & 623.031 & 1.04 & 0.99 \\       
 579 & 1342211121                & 2455544.23532 & R & 663.046 &   1.515 & 160.0 & 650.683 & 623.031 & 1.04 & 0.99 \\       
\noalign{\smallskip}
 789 & 1342223982                & 2455754.55794 & B & 886.706 &   2.181 &  70.0 & 901.124 & 886.077 & 1.02 & 0.93 \\       
 789 & 1342223983                & 2455754.56198 & B & 881.724 &   2.457 &  70.0 & 896.061 & 886.077 & 1.01 & 0.92 \\       
 789 & 1342223985                & 2455754.56862 & G & 898.301 &   2.007 & 100.0 & 905.545 & 884.473 & 1.02 & 0.96 \\       
 789 & 1342223986                & 2455754.57266 & G & 895.208 &   1.795 & 100.0 & 902.427 & 884.473 & 1.02 & 0.96 \\       
 789 & 1342223982                & 2455754.55794 & R & 672.098 &   1.731 & 160.0 & 659.566 & 636.069 & 1.04 & 0.98 \\       
 789 & 1342223983                & 2455754.56198 & R & 674.463 &   1.364 & 160.0 & 661.887 & 636.069 & 1.04 & 0.99 \\       
 789 & 1342223985                & 2455754.56862 & R & 672.173 &   1.744 & 160.0 & 659.640 & 636.069 & 1.04 & 0.98 \\       
 789 & 1342223986                & 2455754.57266 & R & 674.248 &   1.578 & 160.0 & 661.676 & 636.069 & 1.04 & 0.99 \\       
\noalign{\smallskip}
 957 & 1342235629                & 2455922.44924 & B & 851.802 &   2.365 &  70.0 & 865.652 & 856.207 & 1.01 & 0.92 \\       
 957 & 1342235630                & 2455922.45328 & B & 848.719 &   2.329 &  70.0 & 862.519 & 856.207 & 1.01 & 0.92 \\       
 957 & 1342235632                & 2455922.45992 & G & 866.423 &   1.220 & 100.0 & 873.410 & 854.657 & 1.02 & 0.96 \\       
 957 & 1342235633                & 2455922.46396 & G & 863.524 &   1.516 & 100.0 & 870.488 & 854.657 & 1.02 & 0.95 \\       
 957 & 1342235629                & 2455922.44924 & R & 647.260 &   1.633 & 160.0 & 635.191 & 614.627 & 1.03 & 0.98 \\       
 957 & 1342235630                & 2455922.45328 & R & 650.070 &   1.472 & 160.0 & 637.949 & 614.627 & 1.04 & 0.99 \\       
 957 & 1342235632                & 2455922.45992 & R & 647.276 &   1.632 & 160.0 & 635.207 & 614.627 & 1.03 & 0.98 \\       
 957 & 1342235633                & 2455922.46396 & R & 649.832 &   1.485 & 160.0 & 637.715 & 614.627 & 1.04 & 0.99 \\       
\noalign{\smallskip}
1121 & 1342246772                & 2456086.68500 & B & 830.384 &   2.274 &  70.0 & 843.886 & 835.178 & 1.01 & 0.92 \\       
1121 & 1342246773                & 2456086.68904 & B & 828.958 &   2.260 &  70.0 & 842.437 & 835.178 & 1.01 & 0.92 \\       
1121 & 1342246774                & 2456086.69308 & G & 845.147 &   1.872 & 100.0 & 851.963 & 833.666 & 1.02 & 0.96 \\       
1121 & 1342246775                & 2456086.69712 & G & 844.028 &   2.608 & 100.0 & 850.835 & 833.714 & 1.02 & 0.96 \\       
1121 & 1342246772                & 2456086.68500 & R & 630.782 &   1.643 & 160.0 & 619.021 & 599.531 & 1.03 & 0.98 \\       
1121 & 1342246773                & 2456086.68904 & R & 633.665 &   1.283 & 160.0 & 621.850 & 599.531 & 1.04 & 0.98 \\       
1121 & 1342246774                & 2456086.69308 & R & 630.943 &   1.667 & 160.0 & 619.179 & 599.531 & 1.03 & 0.98 \\       
1121 & 1342246775                & 2456086.69712 & R & 634.054 &   1.282 & 160.0 & 622.232 & 599.566 & 1.04 & 0.99 \\       
\noalign{\smallskip}
1310 & 1342257193                & 2456275.57361 & B & 873.132 &   2.372 &  70.0 & 887.329 & 882.711 & 1.01 & 0.92 \\       
1310 & 1342257194                & 2456275.57765 & B & 870.025 &   2.327 &  70.0 & 884.172 & 882.661 & 1.00 & 0.91 \\       
1310 & 1342257195                & 2456275.58169 & G & 891.126 &   2.181 & 100.0 & 898.312 & 881.062 & 1.02 & 0.96 \\       
1310 & 1342257196                & 2456275.58573 & G & 887.978 &   2.013 & 100.0 & 895.139 & 881.062 & 1.02 & 0.95 \\       
1310 & 1342257193                & 2456275.57361 & R & 665.387 &   1.675 & 160.0 & 652.980 & 633.652 & 1.03 & 0.98 \\       
1310 & 1342257194                & 2456275.57765 & R & 667.418 &   1.524 & 160.0 & 654.974 & 633.616 & 1.03 & 0.98 \\       
1310 & 1342257195                & 2456275.58169 & R & 665.408 &   1.677 & 160.0 & 653.001 & 633.616 & 1.03 & 0.98 \\       
1310 & 1342257196                & 2456275.58573 & R & 666.522 &   1.506 & 160.0 & 654.094 & 633.616 & 1.03 & 0.98 \\       
\noalign{\smallskip}\hline
\end{longtable}
\end{longtab}
}

\subsection{Neptune}
\label{sec:neptune}

\begin{figure}[h!tb]
\centering
  \rotatebox{90}{\resizebox{!}{\hsize}{\includegraphics{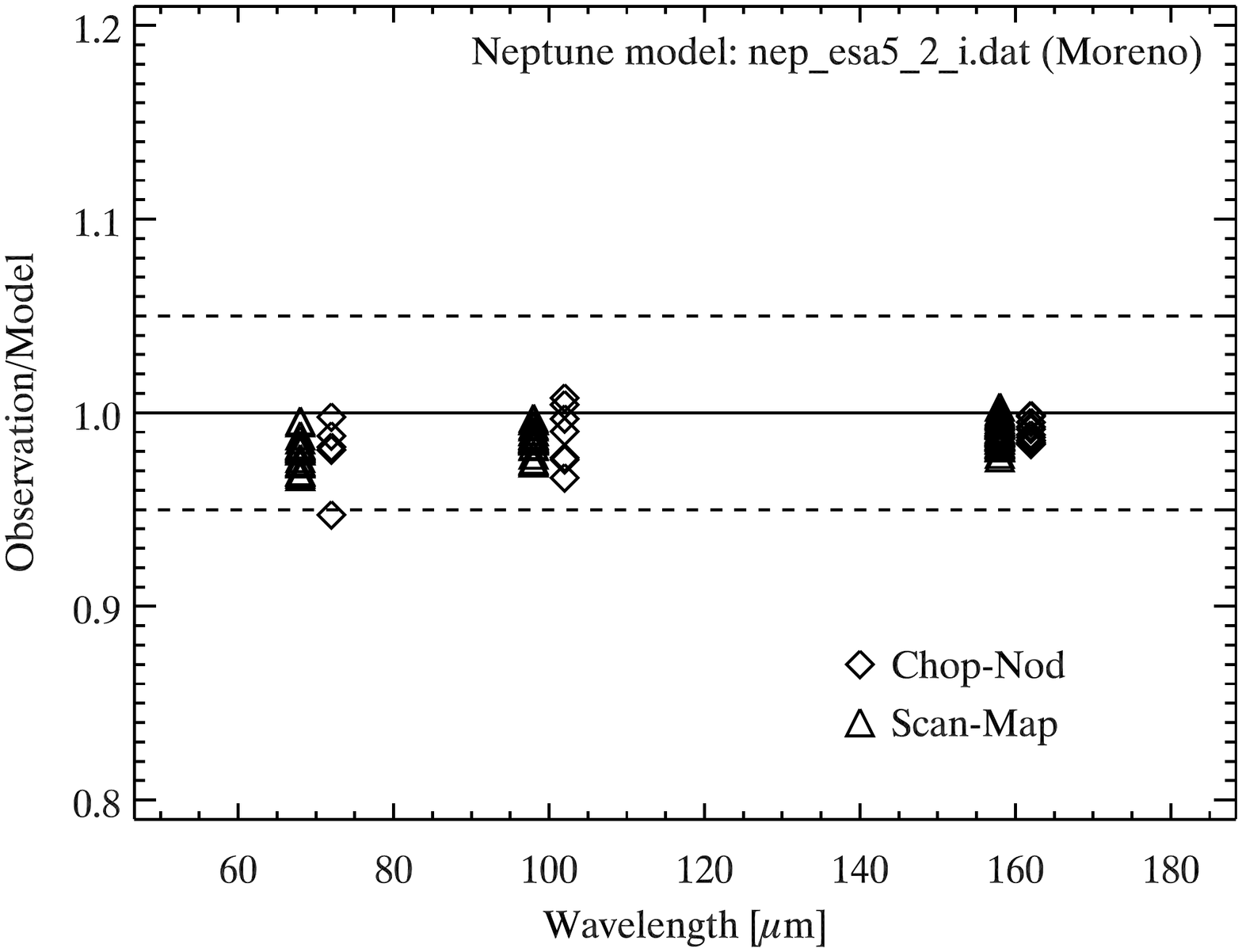}}}
  \rotatebox{90}{\resizebox{!}{\hsize}{\includegraphics{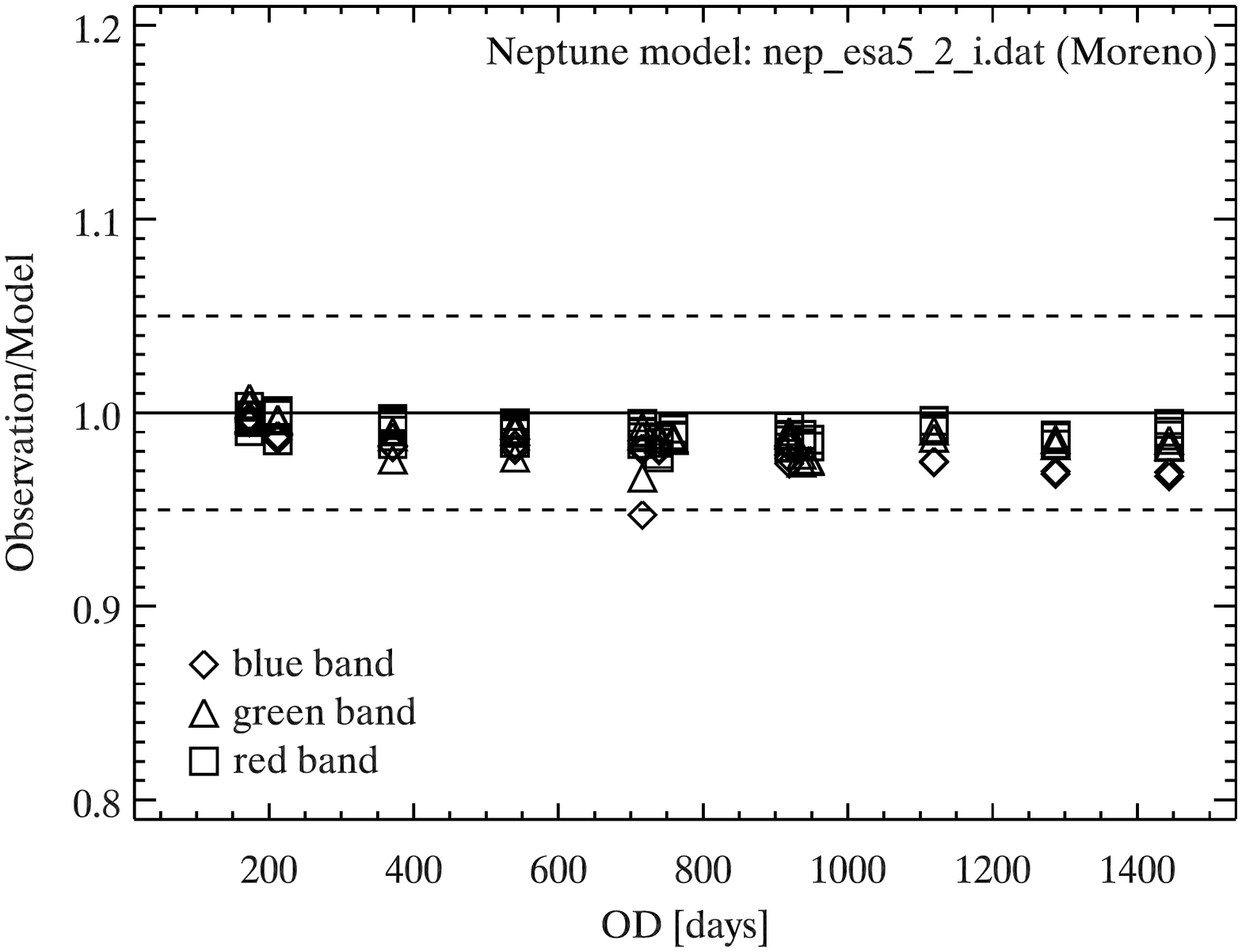}}}
  \caption{All PACS photometer observations of Neptune divided by the
           corresponding model predictions by R.\ Moreno. Top: as a function of wavelength
           (chop-nod and scan-map observations are shown with different symbols).
           Bottom: as a function of OD.
\label{fig:neptune}}
\end{figure}

Table~\ref{tbl:neptune} contains the reduced and calibrated fluxes of 
all Neptune observations obtained with the Herschel-PACS photometer.
Each line is connected to an output map that is related to a single OBSID based either on a
single scan direction or a single chop-nod observation. 
The observing geometry and the relevant
instrument and satellite configuration parameters are given in Tables~\ref{tbl:pacs_chopnod}
and \ref{tbl:pacs_scanmap1}. The absolute calibrated flux densities (column "Flux") are
based on a flat energy spectrum ($\nu$F$_{\nu}$ = const.) with uncertainties
coming from the signal processing and aperture photometry of the final maps.
The color-corrected flux densities at the 
PACS reference wavelength are given in column $FD_{cc}$ , where color-correction factors
from Table~\ref{tbl:cc} were used. The model predictions are based on the
Neptune model "nep\_esa5\_2\_i.dat" (production in 2014), which was taken from the
HCalSG web pages at {\tt ftp://ftp.sciops.esa.int/\-planets/\-originalData/\-esa5/}.
The radiative transfer modeling includes continuum opacities from collision-induced absorption
of H$_2$, He, and CH$_4$. The thermal profiles go back to
Voyager measurements (Pearl \& Conrath \cite{pearl91}, Lindal et al.\ \cite{lindal92})
and fit the IR measurements
of Akari spectra of Neptune's stratosphere (Fletcher et al.\ \cite{fletcher10}).
The Neptune model also fits the line-to-continuum ratio (i.e.,\ relative measurements) of SPIRE
and PACS spectra including CO and HD lines (Moreno et al.\ 2016, in preparation).
The model brightness temperatures were combined with Neptune's angular diameter
as seen from Herschel, taking into account the equatorial
and polar radii and the sub-observer latitude (see footnote in Sect.~\ref{sec:uranus}).
The last column of Table~\ref{tbl:neptune}
contains the calculated ratio between $FD_{cc}$ and the corresponding model prediction
at the given reference wavelength.
The final observation-to-model ratios are shown in Fig.~\ref{fig:neptune}.
The two PACS observing modes - chop-nod and scan-map techniques - are shown with
different symbols and at slightly shifted wavelengths ($\pm$2\,$\mu$m)
for better visibility. The absolute uncertainties for the Neptune model were estimated to be 5\%,
similar to the stellar model uncertainties for the five fiducial stars.

{\small
\begin{longtab}
\begin{longtable}{rlcrrrrlll}
\caption[]{Observational results of the Herschel-PACS photometer observations of Neptune.} \\
\hline\noalign{\smallskip}
   &       &                   & PACS & Flux & Unc.\ & $\lambda_{ref}$ & FD$_{cc}$ & FD$_{model}$ & Ratio \\
OD & OBSID & observ.\ mid-time & Band & [Jy] & [Jy]  & [$\mu$m]        & [Jy]      & [Jy]         & FD$_{cc}$/FD$_{model}$ \\
\noalign{\smallskip}\hline\noalign{\smallskip}
\multicolumn{10}{l}{chop-nod:} \\
\endfirsthead                       
\caption[]{\emph{continued}} \\                                                     
\hline\noalign{\smallskip}                                                                                  
   &       &                   & PACS & Flux & Unc.\ & $\lambda_{ref}$ & FD$_{cc}$ & FD$_{model}$ & Ratio \\
OD & OBSID & observ.\ mid-time & Band & [Jy] & [Jy]  & [$\mu$m]        & [Jy]      & [Jy]         & FD$_{cc}$/FD$_{model}$ \\
\noalign{\smallskip}\hline\noalign{\smallskip}  \endhead                                                    
\hline                                                                                                      
\multicolumn{10}{c}{\emph{continued on next page}} \endfoot                                                 
\noalign{\smallskip}\hline \endlastfoot                                                                     
\label{tbl:neptune}
 173 &   1342186637              & 2455138.56756 & B & 366.939 &   0.612 &  70.0 & 372.905 & 373.735 & 1.00 \\
 173 &   1342186637              & 2455138.56756 & R & 272.863 &   3.096 & 160.0 & 267.513 & 267.860 & 1.00 \\
 173 &   1342186638              & 2455138.57417 & G & 372.860 &   0.580 & 100.0 & 375.488 & 373.938 & 1.00 \\
 173 &   1342186638              & 2455138.57417 & R & 271.858 &   3.087 & 160.0 & 266.527 & 267.860 & 1.00 \\
 173 &   1342186643              & 2455138.58808 & G & 374.152 &   0.573 & 100.0 & 376.790 & 373.938 & 1.01 \\
 173 &   1342186643              & 2455138.58808 & R & 272.697 &   3.160 & 160.0 & 267.350 & 267.860 & 1.00 \\
 212 &   1342188052              & 2455178.10290 & G & 354.153 &   0.734 & 100.0 & 356.650 & 357.790 & 1.00 \\
 212 &   1342188052              & 2455178.10290 & R & 257.683 &   3.417 & 160.0 & 252.630 & 256.293 & 0.99 \\
 212 &   1342188055              & 2455178.11436 & B & 347.654 &   0.074 &  70.0 & 353.307 & 357.596 & 0.99 \\
 212 &   1342188055              & 2455178.11436 & R & 257.644 &   3.425 & 160.0 & 252.592 & 256.293 & 0.99 \\
 371 &   1342196724              & 2455336.47257 & G & 357.696 &   0.693 & 100.0 & 360.218 & 369.190 & 0.98 \\
 371 &   1342196724              & 2455336.47257 & R & 265.381 &   3.622 & 160.0 & 260.177 & 264.459 & 0.98 \\
 540 &   1342209039              & 2455505.54412 & G & 363.158 &   0.617 & 100.0 & 365.718 & 374.431 & 0.98 \\
 540 &   1342209039              & 2455505.54412 & R & 269.358 &   3.027 & 160.0 & 264.076 & 268.213 & 0.98 \\
 540 &   1342209042              & 2455505.55456 & B & 361.160 &   0.590 &  70.0 & 367.033 & 374.227 & 0.98 \\
 540 &   1342209042              & 2455505.55456 & R & 271.239 &   3.081 & 160.0 & 265.921 & 268.213 & 0.99 \\
 716 &   1342220894              & 2455681.48449 & B & 335.675 &   0.624 &  70.0 & 341.133 & 360.137 & 0.95 \\
 716 &   1342220894              & 2455681.48449 & R & 259.927 &   3.487 & 160.0 & 254.830 & 258.114 & 0.99 \\
 716 &   1342220897              & 2455681.49552 & G & 345.814 &   0.606 & 100.0 & 348.252 & 360.365 & 0.97 \\
 716 &   1342220897              & 2455681.49552 & R & 259.015 &   3.333 & 160.0 & 253.936 & 258.137 & 0.98 \\
 919 &   1342232522              & 2455884.64052 & B & 357.180 &   0.577 &  70.0 & 362.988 & 369.544 & 0.98 \\
 919 &   1342232522              & 2455884.64052 & R & 268.174 &   3.133 & 160.0 & 262.916 & 264.856 & 0.99 \\
 919 &   1342232525              & 2455884.65156 & G & 363.611 &   0.625 & 100.0 & 366.174 & 369.745 & 0.99 \\
 919 &   1342232525              & 2455884.65156 & R & 267.067 &   3.274 & 160.0 & 261.830 & 264.856 & 0.99 \\
\noalign{\smallskip}
\multicolumn{10}{l}{scan-map:} \\
 173 &   1342186639              & 2455138.57697 & B & 366.234 &   0.953 &  70.0 & 372.189 & 373.735 & 1.00 \\
 173 &   1342186639              & 2455138.57697 & R & 272.688 &   0.578 & 160.0 & 267.341 & 267.860 & 1.00 \\
 173 &   1342186640              & 2455138.57977 & B & 365.895 &   0.996 &  70.0 & 371.844 & 373.735 & 0.99 \\
 173 &   1342186640              & 2455138.57977 & R & 274.044 &   0.409 & 160.0 & 268.671 & 267.860 & 1.00 \\
 173 &   1342186641              & 2455138.58258 & G & 369.927 &   0.892 & 100.0 & 372.535 & 373.938 & 1.00 \\
 173 &   1342186641              & 2455138.58258 & R & 270.638 &   0.587 & 160.0 & 265.331 & 267.860 & 0.99 \\
 173 &   1342186642              & 2455138.58538 & G & 370.323 &   0.500 & 100.0 & 372.934 & 373.938 & 1.00 \\
 173 &   1342186642              & 2455138.58538 & R & 272.891 &   0.516 & 160.0 & 267.540 & 267.860 & 1.00 \\
 212 &   1342188050              & 2455178.09582 & G & 353.854 &   0.820 & 100.0 & 356.348 & 357.790 & 1.00 \\
 212 &   1342188050              & 2455178.09582 & R & 261.431 &   0.542 & 160.0 & 256.305 & 256.293 & 1.00 \\
 212 &   1342188051              & 2455178.10019 & G & 353.944 &   0.852 & 100.0 & 356.439 & 357.790 & 1.00 \\
 212 &   1342188051              & 2455178.10019 & R & 261.651 &   0.578 & 160.0 & 256.521 & 256.293 & 1.00 \\
 212 &   1342188053              & 2455178.10727 & B & 346.981 &   0.796 &  70.0 & 352.623 & 357.596 & 0.99 \\
 212 &   1342188053              & 2455178.10727 & R & 261.247 &   0.643 & 160.0 & 256.125 & 256.293 & 1.00 \\
 212 &   1342188054              & 2455178.11165 & B & 347.727 &   0.816 &  70.0 & 353.381 & 357.596 & 0.99 \\
 212 &   1342188054              & 2455178.11165 & R & 261.144 &   0.583 & 160.0 & 256.024 & 256.293 & 1.00 \\
 371 &   1342196725              & 2455336.47648 & G & 363.033 &   0.828 & 100.0 & 365.592 & 369.190 & 0.99 \\
 371 &   1342196725              & 2455336.47648 & R & 267.308 &   0.569 & 160.0 & 262.067 & 264.459 & 0.99 \\
 371 &   1342196726              & 2455336.48041 & G & 362.101 &   0.844 & 100.0 & 364.654 & 369.190 & 0.99 \\
 371 &   1342196726              & 2455336.48041 & R & 268.028 &   0.649 & 160.0 & 262.773 & 264.459 & 0.99 \\
 371 &   1342196727              & 2455336.48433 & B & 356.769 &   0.784 &  70.0 & 362.570 & 368.989 & 0.98 \\
 371 &   1342196727              & 2455336.48433 & R & 268.451 &   0.667 & 160.0 & 263.187 & 264.459 & 1.00 \\
 371 &   1342196728              & 2455336.48825 & B & 356.584 &   0.805 &  70.0 & 362.382 & 368.989 & 0.98 \\
 371 &   1342196728              & 2455336.48825 & R & 268.881 &   0.658 & 160.0 & 263.609 & 264.459 & 1.00 \\
 540 &   1342209040              & 2455505.54803 & G & 369.456 &   0.834 & 100.0 & 372.060 & 374.431 & 0.99 \\
 540 &   1342209040              & 2455505.54803 & R & 270.919 &   0.671 & 160.0 & 265.607 & 268.213 & 0.99 \\
 540 &   1342209041              & 2455505.55196 & G & 368.218 &   0.483 & 100.0 & 370.814 & 374.431 & 0.99 \\
 540 &   1342209041              & 2455505.55196 & R & 270.729 &   0.588 & 160.0 & 265.421 & 268.213 & 0.99 \\
 540 &   1342209043              & 2455505.55848 & B & 362.158 &   0.786 &  70.0 & 368.047 & 374.227 & 0.98 \\
 540 &   1342209043              & 2455505.55848 & R & 272.120 &   0.671 & 160.0 & 266.784 & 268.213 & 0.99 \\
 540 &   1342209044              & 2455505.56241 & B & 362.000 &   0.656 &  70.0 & 367.886 & 374.227 & 0.98 \\
 540 &   1342209044              & 2455505.56241 & R & 271.890 &   0.602 & 160.0 & 266.559 & 268.213 & 0.99 \\
 716 &   1342220895              & 2455681.48888 & B & 347.936 &   0.763 &  70.0 & 353.594 & 360.169 & 0.98 \\
 716 &   1342220895              & 2455681.48888 & R & 261.776 &   0.636 & 160.0 & 256.643 & 258.137 & 0.99 \\
 716 &   1342220896              & 2455681.49292 & B & 347.487 &   0.775 &  70.0 & 353.137 & 360.169 & 0.98 \\
 716 &   1342220896              & 2455681.49292 & R & 261.817 &   0.576 & 160.0 & 256.683 & 258.137 & 0.99 \\
 716 &   1342220898              & 2455681.49955 & G & 355.286 &   0.810 & 100.0 & 357.791 & 360.365 & 0.99 \\
 716 &   1342220898              & 2455681.49955 & R & 260.732 &   0.783 & 160.0 & 255.620 & 258.137 & 0.99 \\
 716 &   1342220899              & 2455681.50359 & G & 353.601 &   0.587 & 100.0 & 356.094 & 360.365 & 0.99 \\
 716 &   1342220899              & 2455681.50359 & R & 260.569 &   0.562 & 160.0 & 255.460 & 258.137 & 0.99 \\
 739 &   1342221604              & 2455705.06015 & B & 356.927 &   0.789 &  70.0 & 362.731 & 369.609 & 0.98 \\
 739 &   1342221604              & 2455705.06015 & R & 266.494 &   0.661 & 160.0 & 261.269 & 264.903 & 0.99 \\
 739 &   1342221605              & 2455705.06713 & B & 356.592 &   0.803 &  70.0 & 362.390 & 369.609 & 0.98 \\
 739 &   1342221605              & 2455705.06713 & R & 266.023 &   0.591 & 160.0 & 260.807 & 264.903 & 0.98 \\
 739 &   1342221606              & 2455705.07116 & G & 362.998 &   0.819 & 100.0 & 365.557 & 369.811 & 0.99 \\
 739 &   1342221606              & 2455705.07116 & R & 263.938 &   0.655 & 160.0 & 258.763 & 264.903 & 0.98 \\
 739 &   1342221607              & 2455705.07513 & G & 362.340 &   0.846 & 100.0 & 364.894 & 369.843 & 0.99 \\
 739 &   1342221607              & 2455705.07513 & R & 264.450 &   0.587 & 160.0 & 259.265 & 264.926 & 0.98 \\
 759 &   1342222561              & 2455724.68860 & G & 370.019 &   0.975 & 100.0 & 372.627 & 377.989 & 0.99 \\
 759 &   1342222561              & 2455724.68860 & R & 273.894 &   0.603 & 160.0 & 268.524 & 270.762 & 0.99 \\
 759 &   1342222562              & 2455724.72412 & G & 370.429 &   0.586 & 100.0 & 373.040 & 377.989 & 0.99 \\
 759 &   1342222562              & 2455724.72412 & R & 272.961 &   0.527 & 160.0 & 267.609 & 270.762 & 0.99 \\
 759 &   1342222563              & 2455724.75953 & G & 370.223 &   0.979 & 100.0 & 372.833 & 377.989 & 0.99 \\
 759 &   1342222563              & 2455724.75953 & R & 273.838 &   0.598 & 160.0 & 268.469 & 270.762 & 0.99 \\
 759 &   1342222564              & 2455724.79493 & G & 370.573 &   0.583 & 100.0 & 373.185 & 378.022 & 0.99 \\
 759 &   1342222564              & 2455724.79493 & R & 272.766 &   0.524 & 160.0 & 267.418 & 270.785 & 0.99 \\
 919 &   1342232523              & 2455884.64491 & B & 355.081 &   0.780 &  70.0 & 360.855 & 369.544 & 0.98 \\
 919 &   1342232523              & 2455884.64491 & R & 266.884 &   0.654 & 160.0 & 261.651 & 264.856 & 0.99 \\
 919 &   1342232524              & 2455884.64895 & B & 354.065 &   0.788 &  70.0 & 359.822 & 369.544 & 0.97 \\
 919 &   1342232524              & 2455884.64895 & R & 266.345 &   0.594 & 160.0 & 261.123 & 264.856 & 0.99 \\
 919 &   1342232526              & 2455884.65558 & G & 362.990 &   0.829 & 100.0 & 365.549 & 369.745 & 0.99 \\
 919 &   1342232526              & 2455884.65558 & R & 265.308 &   0.660 & 160.0 & 260.106 & 264.856 & 0.98 \\
 919 &   1342232527              & 2455884.65962 & G & 361.835 &   0.607 & 100.0 & 364.386 & 369.745 & 0.99 \\
 919 &   1342232527              & 2455884.65962 & R & 265.327 &   0.583 & 160.0 & 260.125 & 264.856 & 0.98 \\
 936 &   1342234207              & 2455901.05921 & G & 351.078 &   0.513 & 100.0 & 353.553 & 362.917 & 0.97 \\
 936 &   1342234207              & 2455901.05921 & R & 260.956 &   0.514 & 160.0 & 255.839 & 259.965 & 0.98 \\
 936 &   1342234208              & 2455901.09703 & G & 352.495 &   0.443 & 100.0 & 354.980 & 362.885 & 0.98 \\
 936 &   1342234208              & 2455901.09703 & R & 262.127 &   0.575 & 160.0 & 256.987 & 259.942 & 0.99 \\
 947 &   1342234435              & 2455912.34192 & G & 347.136 &   0.534 & 100.0 & 349.583 & 358.562 & 0.97 \\
 947 &   1342234435              & 2455912.34192 & R & 257.426 &   0.661 & 160.0 & 252.378 & 256.845 & 0.98 \\
 947 &   1342234436              & 2455912.37884 & G & 347.371 &   0.471 & 100.0 & 349.820 & 358.562 & 0.98 \\
 947 &   1342234436              & 2455912.37884 & R & 258.321 &   0.563 & 160.0 & 253.256 & 256.845 & 0.99 \\
1119 &   1342246671              & 2456084.58171 & G & 368.955 &   0.842 & 100.0 & 371.556 & 375.056 & 0.99 \\
1119 &   1342246671              & 2456084.58171 & R & 271.805 &   0.671 & 160.0 & 266.475 & 268.660 & 0.99 \\
1119 &   1342246672              & 2456084.59417 & G & 367.594 &   0.487 & 100.0 & 370.185 & 375.056 & 0.99 \\
1119 &   1342246672              & 2456084.59417 & R & 271.487 &   0.601 & 160.0 & 266.164 & 268.660 & 0.99 \\
1119 &   1342246673              & 2456084.59822 & B & 359.597 &   0.793 &  70.0 & 365.444 & 374.852 & 0.97 \\
1119 &   1342246673              & 2456084.59822 & R & 272.899 &   0.677 & 160.0 & 267.548 & 268.660 & 1.00 \\
1119 &   1342246674              & 2456084.60226 & B & 359.418 &   0.737 &  70.0 & 365.262 & 374.852 & 0.97 \\
1119 &   1342246674              & 2456084.60226 & R & 272.848 &   0.606 & 160.0 & 267.498 & 268.660 & 1.00 \\
1287 &   1342255709              & 2456252.87637 & G & 362.095 &   0.828 & 100.0 & 364.648 & 369.713 & 0.99 \\
1287 &   1342255709              & 2456252.87637 & R & 265.727 &   0.652 & 160.0 & 260.517 & 264.833 & 0.98 \\
1287 &   1342255710              & 2456252.88317 & G & 360.922 &   0.597 & 100.0 & 363.466 & 369.713 & 0.98 \\
1287 &   1342255710              & 2456252.88317 & R & 265.704 &   0.594 & 160.0 & 260.494 & 264.833 & 0.98 \\
1287 &   1342255711              & 2456252.88729 & B & 352.586 &   0.780 &  70.0 & 358.319 & 369.511 & 0.97 \\
1287 &   1342255711              & 2456252.88729 & R & 266.719 &   0.651 & 160.0 & 261.489 & 264.833 & 0.99 \\
1287 &   1342255712              & 2456252.89133 & B & 352.057 &   0.719 &  70.0 & 357.781 & 369.479 & 0.97 \\
1287 &   1342255712              & 2456252.89133 & R & 266.919 &   0.588 & 160.0 & 261.685 & 264.809 & 0.99 \\
1444 &   1342270939              & 2456409.23071 & G & 350.582 &   0.667 & 100.0 & 353.053 & 358.112 & 0.99 \\
1444 &   1342270939              & 2456409.23071 & R & 256.998 &   0.629 & 160.0 & 251.959 & 256.523 & 0.98 \\
1444 &   1342270940              & 2456409.23766 & G & 349.436 &   0.580 & 100.0 & 351.899 & 358.112 & 0.98 \\
1444 &   1342270940              & 2456409.23766 & R & 259.081 &   0.582 & 160.0 & 254.001 & 256.523 & 0.99 \\
1444 &   1342270941              & 2456409.24170 & B & 341.380 &   0.681 &  70.0 & 346.931 & 357.917 & 0.97 \\
1444 &   1342270941              & 2456409.24170 & R & 258.577 &   0.626 & 160.0 & 253.507 & 256.523 & 0.99 \\
1444 &   1342270942              & 2456409.24574 & B & 340.622 &   0.623 &  70.0 & 346.161 & 357.917 & 0.97 \\
1444 &   1342270942              & 2456409.24574 & R & 260.128 &   0.578 & 160.0 & 255.027 & 256.523 & 0.99 \\
\noalign{\smallskip}\hline
\end{longtable}
\end{longtab}
}

\subsection{Callisto, Ganymede, and Titan}

\begin{figure}[h!tb]
\centering
  \rotatebox{0}{\resizebox{\hsize}{!}{\includegraphics{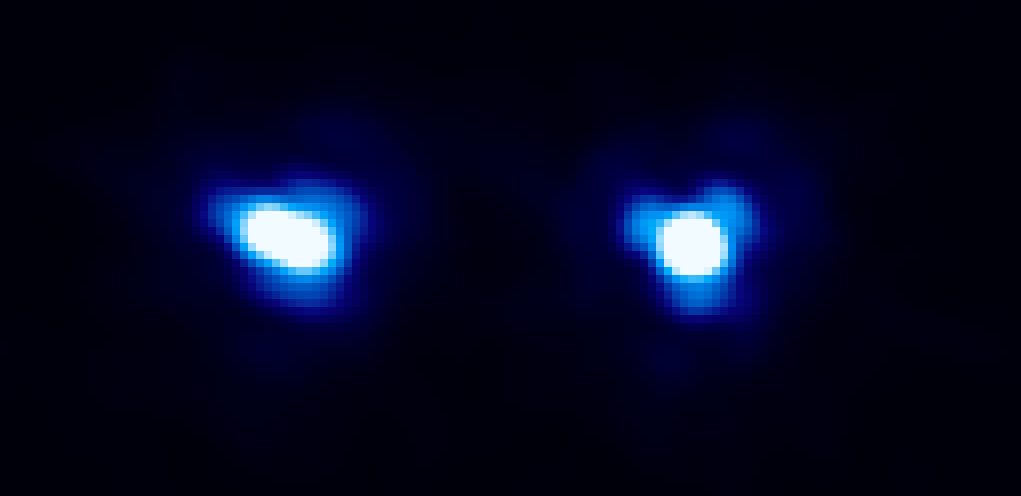}}}

  \vspace*{0.1cm}

  \rotatebox{0}{\resizebox{\hsize}{!}{\includegraphics{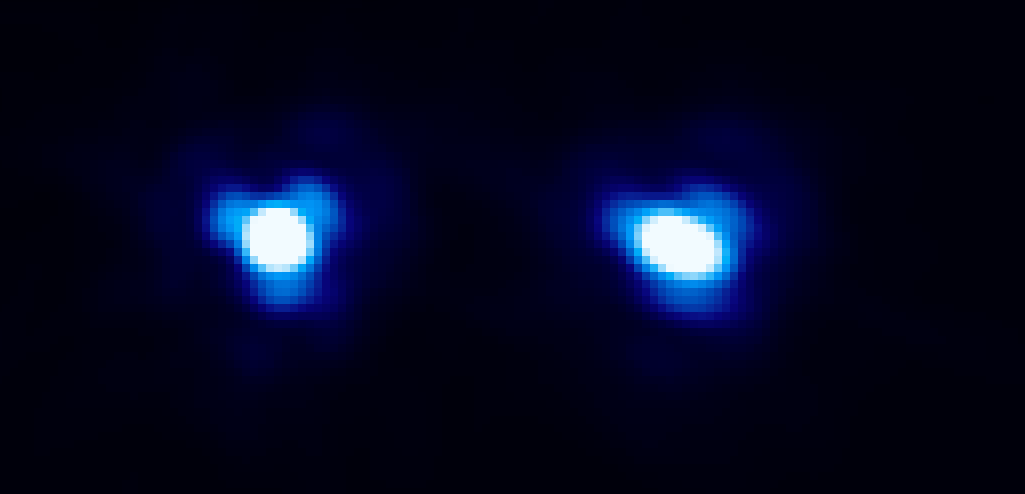}}}
  \caption{Combined Callisto and Ganymede measurement with the PACS
           photometer at 70\,$\mu$m. The two sources were separated by
           about 55$^{\prime \prime}$ and moved with different apparent
           motions (see Table~\ref{tbl:call_gany_titan}) at the time
           of the observations.
           Top: PACS observing frames centered on Callisto (right source). Bottom:
           centered on Ganymede (left source). The angular separation is about 6$^{\prime}$
           between Jupiter and Callisto and about 5$^{\prime}$ between Jupiter
           and Ganymede, with an apparent angular diameter of Jupiter of 40.7$^{\prime \prime}$.
\label{fig:cal_gany}}
\end{figure}

\begin{figure*}[h!tb]
\centering
  \rotatebox{0}{\resizebox{!}{5cm}{\includegraphics{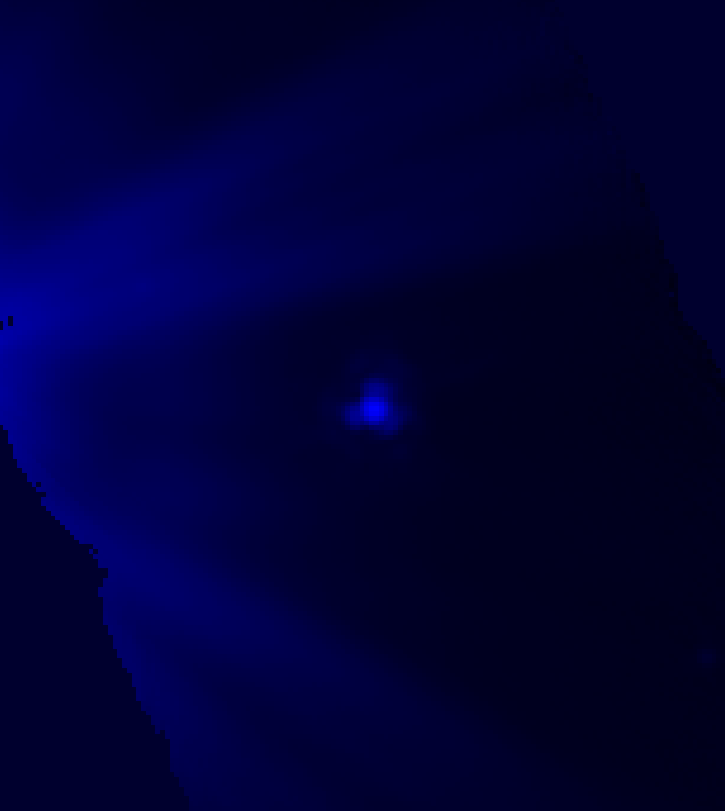}}}
  \rotatebox{0}{\resizebox{!}{5cm}{\includegraphics{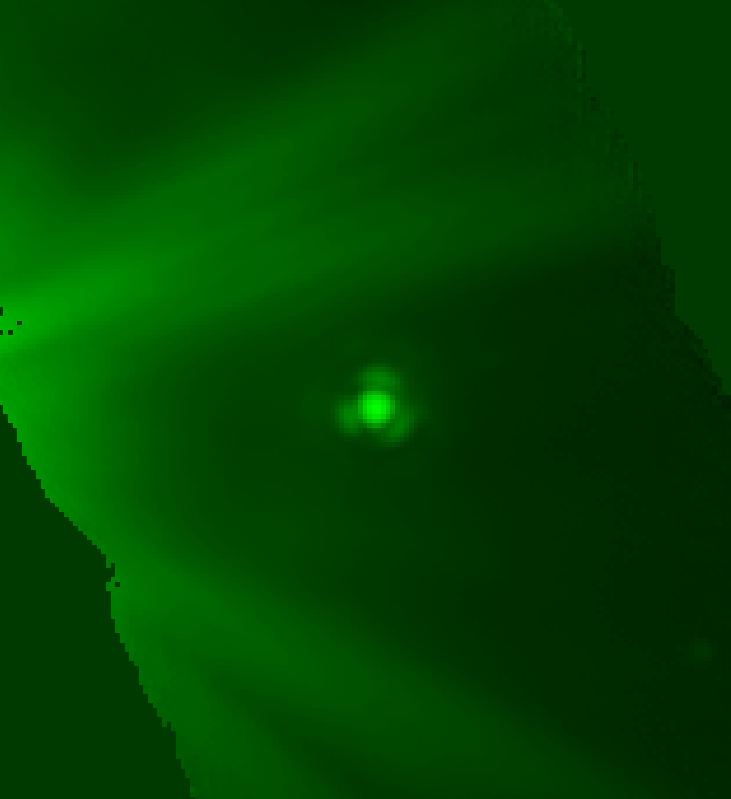}}}
  \rotatebox{0}{\resizebox{!}{5cm}{\includegraphics{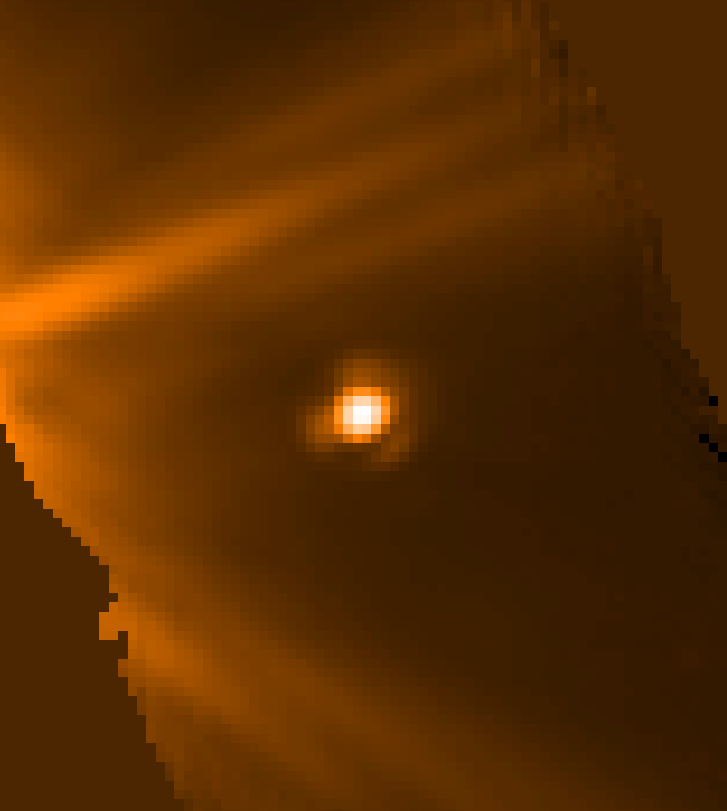}}}
  \caption{Titan measurement with the PACS photometer at 70, 100, and 160\,$\mu$m.
           Titan was located about 3\,arcmin from Saturn. Saturn's PSF wings are
           visible, but did not influence the photometry significantly.
\label{fig:titan_bgr}}
\end{figure*}

The two point-like Jupiter moons Callisto and Ganymede are so bright
(each $>$ 1000\,Jy at 70\,$\mu$m) that they allow instantaneous determination
of their brightness peaks within the time resolution of a bolometer frame (10\,Hz).
The scan-map observations from OD~981 were used to determine the relative
positions of the bolometer matrix pixels, the size of the matrix gaps, rotation
angles of the matrices, and distortion effects. 
The measurements were performed when the two moons were at the farthest possible
distance from Jupiter (Callisto at 351-361$^{\prime \prime}$
and Ganymede at 303-304$^{\prime \prime}$
from Jupiter's body center, with Jupiter having an apparent Herschel-centric
size of 40.7$^{\prime \prime}$).

Titan, the largest moon of Saturn, was observed as part of the PACS photometer
flux calibration program in the context of linearity checks and cross-calibration
in the intermediate flux regime. It was observed when it was at the largest separation from
Saturn at 183$^{\prime \prime}$, with Saturn having an apparent Herschel-centric
size of 17.7$^{\prime \prime}$ (see Fig.~\ref{fig:titan_bgr}).

Table~\ref{tbl:call_gany_titan} summarises the
apparent motions, sizes, and angular separations during the PACS observations.
The Callisto and Ganymede observations were executed in orthogonal scan directions
of 0$^{\circ}$ and 90$^{\circ}$ with respect to the array orientation
to cope with possible timing problems. The measurements
were also taken in support of pointing jitter assessment during scans and 
have 4\,Hz star tracker telemetry available. The Titan measurements were taken
in standard scan- and cross-scan observations.
 
The model of Titan ({\tt tit\_esa3\_2\_i.dat}\footnote{available from
{\tt ftp://ftp.sciops.esa.int/\-planets/\-originalData/\-esa3/} and soon from the
Herschel Explanatory Library Listings (HELL) at {\tt http://www.cosmos.esa.int/\-web/\-herschel}}
by Moreno) was computed using the radiative transfer model for a disk-averaged geometry,
described in Courtin et al.\ (\cite{courtin11}) and Moreno et al.\ (\cite{moreno12}). This
model includes continuum opacities from collision-induced absorption of N$_2$-CH$_4$ pairs and uses
the thermal profile from Huygens probe measurements (Fulchignoni et al.\ \cite{fulchignoni05})
combined with CIRS measurements (Vinatier et al.\ \cite{vinatier10}). Molecular lines of CO,
HCN, and CH$_4$ and their isotopes are included, as are the HIFI and SPIRE observations.
The final absolute accuracy is better than 5\% for Titan.

The disk-averaged models of Callisto ({\tt call\_esa2\_2\_i.dat}\footnote{available from 
{\tt ftp://ftp.sciops.esa.int/\-planets/\-originalData/\-esa2/}}
by Moreno) and Ganymede ({\tt gany\_esa2\_2\_i.dat}\footnote{available from
{\tt ftp://ftp.sciops.esa.int/\-planets/\-originalData/\-esa2/}}
by Moreno) are based on thermal models of the sub-surface
computed upon the Spencer et al.\ (\cite{spencer89}) algorithm. The models solve the
heat diffusion equation in the planetary surface material as a function of longitude,
latitude, and depth. The thermal inertia used was originally derived by Spencer (\cite{spencer87})
from two-layer models and based on 10-20\,$\mu$m data from Voyager. These models also include the
surface dielectric constant and roughness,
which are fit from the ground-based measurements at mm-wavelength performed with
the IRAM-PdBI and the SMA (Moreno et al.\ \cite{moreno08}).
The model accuracies were estimated to be better than 7\% for Ganymede and Callisto.
Effectively, Callisto and Ganymede continuum models are
not very accurate in the PACS wavelength range. This is probably
linked to the effective albedo and different thermal inertia layers
in the sub-surface, which are very difficult to constrain from disk-averaged
observations.

\begin{figure}[h!tb]
\centering
  \rotatebox{90}{\resizebox{!}{\hsize}{\includegraphics{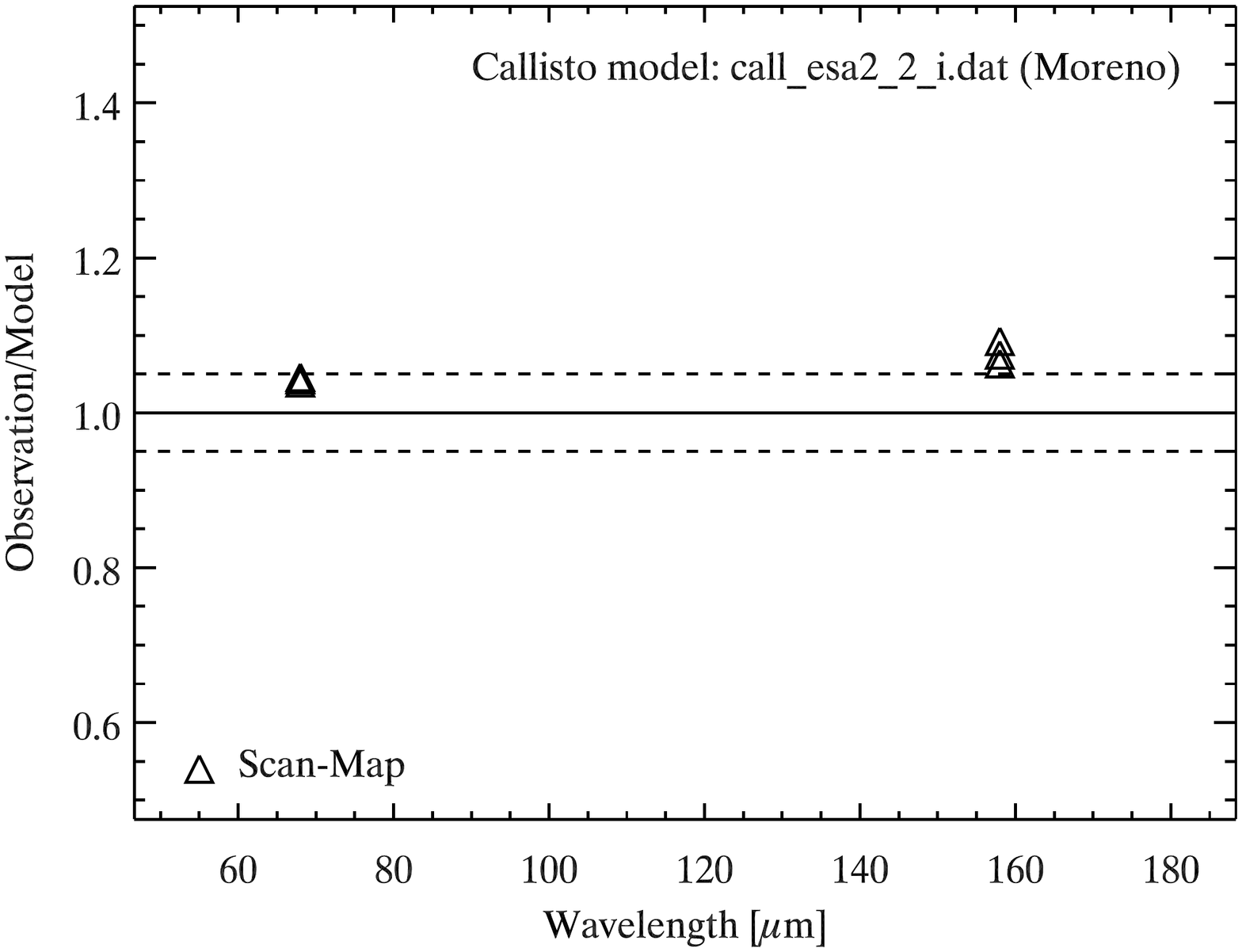}}}
  \rotatebox{90}{\resizebox{!}{\hsize}{\includegraphics{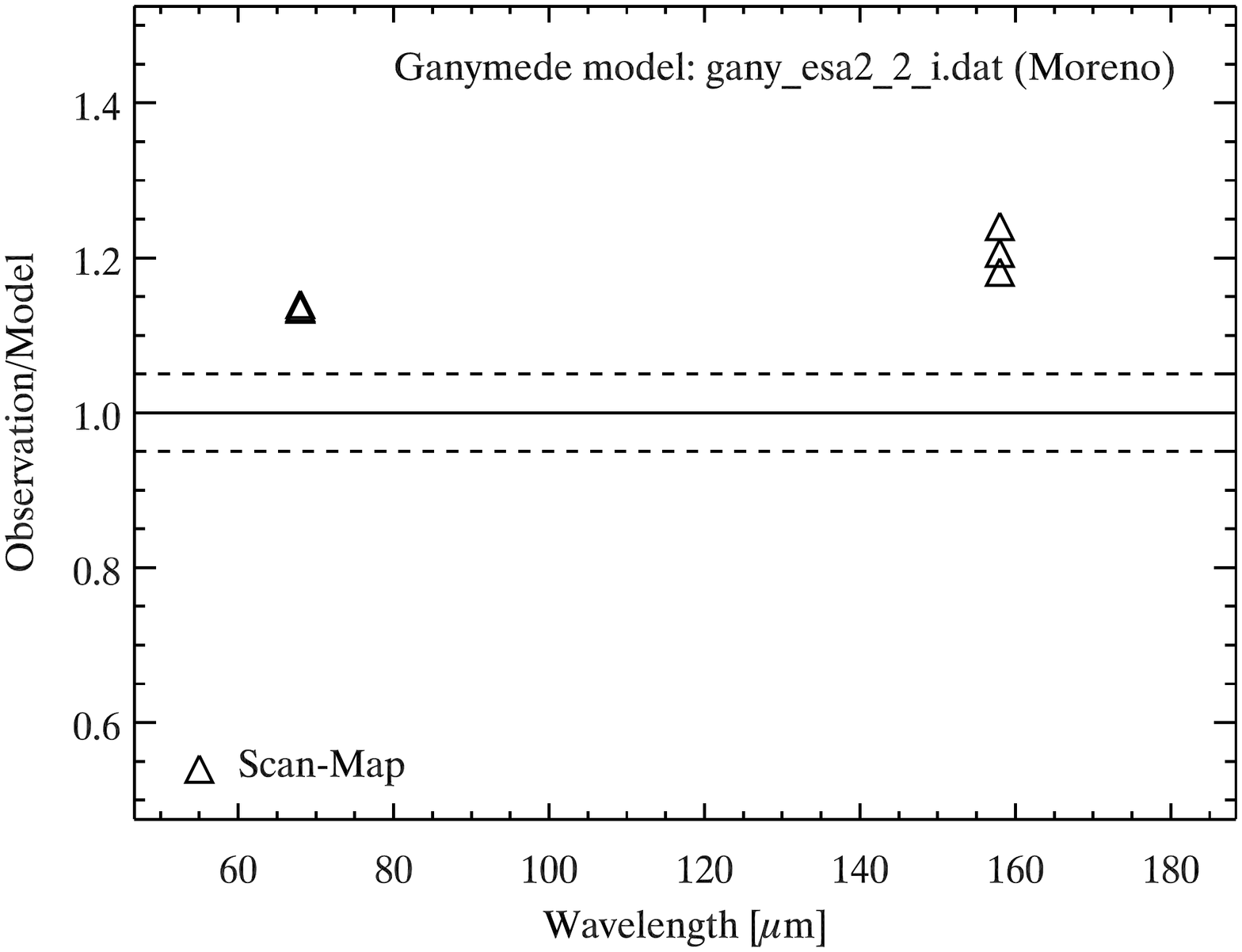}}}
  \rotatebox{90}{\resizebox{!}{\hsize}{\includegraphics{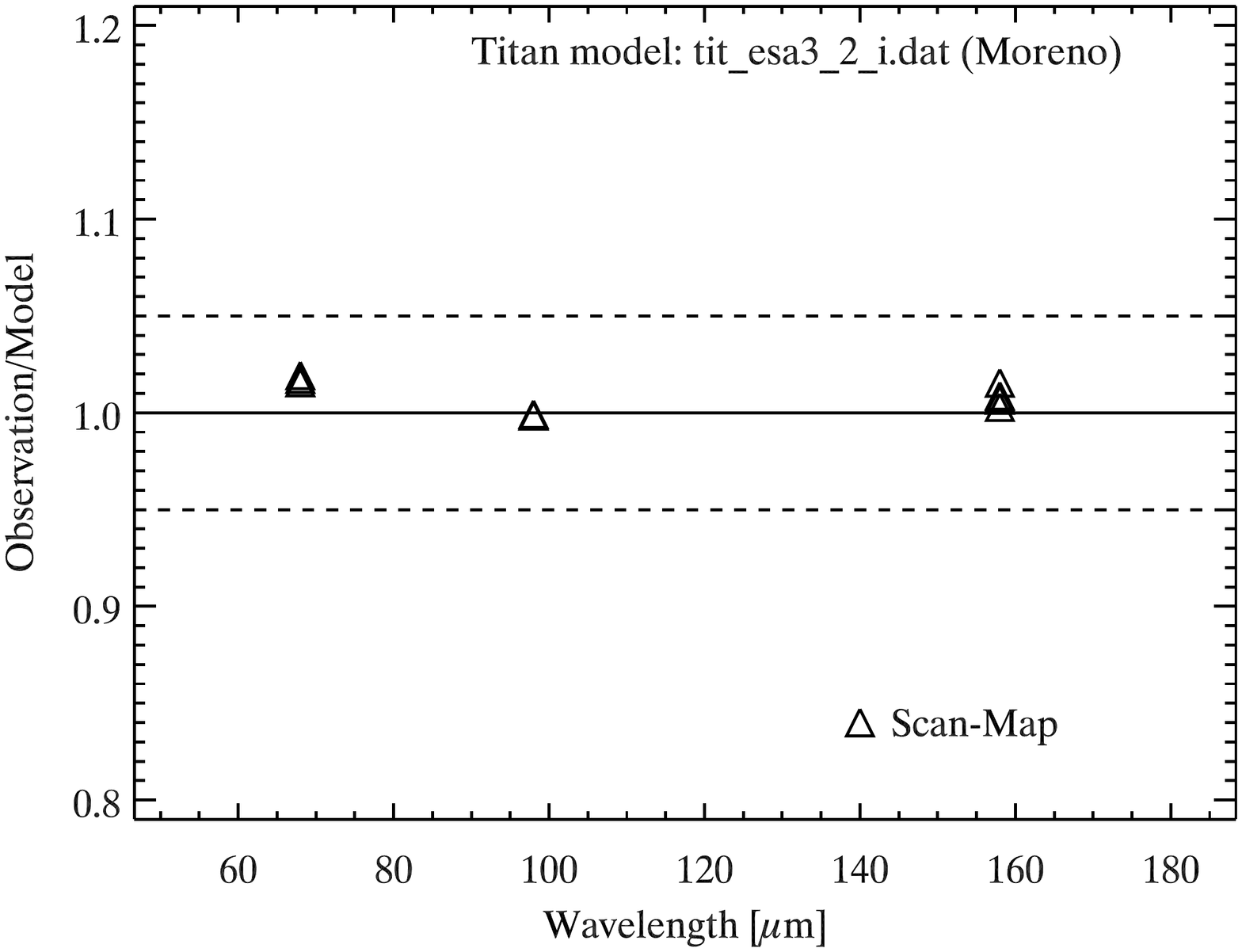}}}
  \caption{All PACS photometer observations of Callisto, Ganymede, and Titan, divided by the
           corresponding model predictions and as a function of wavelength.
           Top: Callisto. Middle: Ganymede. Bottom: Titan.
\label{fig:callisto_ganymede_titan}}
\end{figure}

The model brightness temperatures were combined with the angular diameters (as seen by Herschel
and as listed by JPL/Horizon system under "Ang-diam") of these three satellites and calculated
for the precise observing epoch.
The penultimate columns of Tables~\ref{tbl:callisto}, \ref{tbl:ganymede}, and \ref{tbl:titan}
contain the calculated ratio between $FD_{cc}$ and the corresponding model prediction
at the given reference wavelength.
The final observation-to-model ratios are shown in Fig.~\ref{fig:callisto_ganymede_titan}.

\begin{table*}
{\small
\caption[]{Observational results of the Herschel-PACS photometer observations of Callisto. Columns are
           described in the text, the brightness temperatures T$_b^{obs}$ were calculated from the
           observed fluxes, the apparent effective size, and the total absolute flux uncertainties (see Sect.~\ref{sec:results}).}
\label{tbl:callisto}
\begin{tabular}{rlcrrrrrrll}
\hline\noalign{\smallskip}
   &          &                   & PACS & Flux & Unc.\ & $\lambda_{ref}$ & FD$_{cc}$ & FD$_{model}$ & Ratio                  &  T$_b^{obs}$ \\
OD & OBSID(s) & observ.\ mid-time & Band & [Jy] & [Jy]  & [$\mu$m]        & [Jy]      & [Jy]         & FD$_{cc}$/FD$_{model}$ &  [K]   \\
\noalign{\smallskip}\hline\noalign{\smallskip}
\multicolumn{11}{l}{scan-map:} \\
 981 & 1342238042                & 2455946.74716 & B & 1324.792 &    3.772 &  70.0 & 1327.447 & 1272.926 & 1.04 & 147.7 $\pm$ 5.6 \\
 981 & 1342238042                & 2455946.74716 & R &  416.483 &    3.289 & 160.0 &  394.771 &  371.255 & 1.06 & 146.0 $\pm$ 6.5 \\
 981 & 1342238043                & 2455946.77907 & B & 1328.038 &    8.141 &  70.0 & 1330.699 & 1272.554 & 1.05 & 147.8 $\pm$ 5.6 \\
 981 & 1342238043                & 2455946.77907 & R &  427.658 &   12.807 & 160.0 &  405.363 &  371.146 & 1.09 & 148.9 $\pm$ 6.7 \\
\noalign{\smallskip}\hline
\end{tabular}
}
\end{table*}

\begin{table*}
{\small
\caption[]{Observational results of the Herschel-PACS photometer observations of Ganymede.}
\label{tbl:ganymede}
\begin{tabular}{rlcrrrrrrll}
\hline\noalign{\smallskip}
   &          &                   & PACS & Flux & Unc.\ & $\lambda_{ref}$ & FD$_{cc}$ & FD$_{model}$ & Ratio                  &  T$_b^{obs}$ \\
OD & OBSID(s) & observ.\ mid-time & Band & [Jy] & [Jy]  & [$\mu$m]        & [Jy]      & [Jy]         & FD$_{cc}$/FD$_{model}$ &  [K] \\
\noalign{\smallskip}\hline\noalign{\smallskip}
\multicolumn{11}{l}{scan-map:} \\
 981 & 1342238042                & 2455946.74716 & B & 1321.756 &    4.899 &  70.0 & 1327.064 & 1168.363 & 1.14 & 134.3 $\pm$ 4.8 \\
 981 & 1342238042                & 2455946.74716 & R &  442.907 &    4.670 & 160.0 &  420.614 &  355.936 & 1.18 & 134.0 $\pm$ 5.8 \\
 981 & 1342238043                & 2455946.77907 & B & 1326.010 &    7.710 &  70.0 & 1331.335 & 1167.895 & 1.14 & 134.5 $\pm$ 4.8 \\
 981 & 1342238043                & 2455946.77907 & R &  464.826 &   15.927 & 160.0 &  441.430 &  355.793 & 1.24 & 138.8 $\pm$ 6.1 \\
\noalign{\smallskip}\hline
\end{tabular}
}
\end{table*}

\begin{table*}
{\small
\caption[]{Observational results of the Herschel-PACS photometer observations of Titan.}
\label{tbl:titan}
\begin{tabular}{rlcrrrrrrll}
\hline\noalign{\smallskip}
   &          &                   & PACS & Flux & Unc.\ & $\lambda_{ref}$ & FD$_{cc}$ & FD$_{model}$ & Ratio                  &  T$_b^{obs}$ \\
OD & OBSID(s) & observ.\ mid-time & Band & [Jy] & [Jy]  & [$\mu$m]        & [Jy]      & [Jy]         & FD$_{cc}$/FD$_{model}$ &  [K]  \\
\noalign{\smallskip}\hline\noalign{\smallskip}
\multicolumn{11}{l}{scan-map:} \\
1138 & 1342247418                & 2456104.14470 & B &   89.380 &    0.656 &  70.0 &   90.741 &   89.029 & 1.02 & 76.8 $\pm$ 1.6 \\
1138 & 1342247419                & 2456104.14470 & B &   89.066 &    0.530 &  70.0 &   90.422 &   89.029 & 1.02 & 76.7 $\pm$ 1.6 \\
1138 & 1342247420                & 2456104.15278 & G &   81.484 &    0.583 & 100.0 &   81.321 &   81.451 & 1.00 & 79.0 $\pm$ 2.2 \\
1138 & 1342247421                & 2456104.15278 & G &   81.538 &    0.593 & 100.0 &   81.375 &   81.451 & 1.00 & 79.0 $\pm$ 2.2 \\
1138 & 1342247418                & 2456104.14874 & R &   53.901 &    2.810 & 160.0 &   52.433 &   51.981 & 1.01 & 82.8 $\pm$ 3.0 \\
1138 & 1342247419                & 2456104.14874 & R &   53.788 &    2.955 & 160.0 &   52.323 &   51.981 & 1.01 & 82.7 $\pm$ 3.0 \\
1138 & 1342247420                & 2456104.14874 & R &   53.853 &    2.612 & 160.0 &   52.386 &   51.981 & 1.01 & 82.8 $\pm$ 3.0 \\
1138 & 1342247421                & 2456104.14874 & R &   54.259 &    2.975 & 160.0 &   52.781 &   51.981 & 1.02 & 83.1 $\pm$ 3.0 \\
\noalign{\smallskip}\hline
\end{tabular}
}
\end{table*}

\begin{table*}
\caption{Summary of the motions, sizes, and
         separations of Callisto, Ganymede, Titan, Jupiter, and
Saturn at start and end times of the PACS observations.
\label{tbl:call_gany_titan}}
\begin{tabular}{llrrrr}
\noalign{\smallskip}\hline\hline\noalign{\smallskip}
       &       & dRA*cosD/dt\tablefootmark{1} & d(DEC)/dt\tablefootmark{1} & ang-sep\tablefootmark{2} & Ang-diam\tablefootmark{3} \\
Object & Time  & [$^{\prime \prime}$/h] & [$^{\prime \prime}$/h]        & [$^{\prime \prime}$] & [$^{\prime \prime}$] \\
\noalign{\smallskip}\hline\noalign{\smallskip}
Callisto  & 2012 Jan 20 05:30 &  17.21 & 7.73 & 361.0 &  1.37 \\
          & 2012 Jan 20 07:00 &  17.36 & 7.78 & 351.6 &  1.37 \\
\noalign{\smallskip}
Ganymede  & 2012 Jan 20 05:30 &  11.68 & 5.87 & 304.2 &  1.50 \\
          & 2012 Jan 20 07:00 &  12.28 & 6.10 & 303.1 &  1.50 \\
\noalign{\smallskip}
Jupiter   & 2012 Jan 20 05:30 &  11.48 & 5.18 & ---   & 40.73 \\
          & 2012 Jan 20 07:00 &  11.51 & 5.19 & ---   & 40.72 \\
\noalign{\smallskip}\hline\noalign{\smallskip}
Titan     & 2012 Jun 25 15:22 &  -0.27 & 0.05 & 183.1 &  0.758 \\
          & 2012 Jun 25 15:45 &  -0.25 & 0.05 & 183.1 &  0.758 \\
\noalign{\smallskip}
Saturn    & 2012 Jun 25 15:22 &  -0.18 & -0.57 & ---   & 17.74 \\
          & 2012 Jun 25 15:45 &  -0.18 & -0.57 & ---   & 17.74 \\
\noalign{\smallskip}\hline
\end{tabular}
\tablefoot{\tablefoottext{1}{The change rate of the target center apparent RA and DEC (airless).
                             d(RA)/dt is multiplied by the cosine of the declination};
           \tablefoottext{2}{target-primary angular separation. The angle between the
                             center of target object and the center of the primary body it
                             revolves around, as seen by the observer};
           \tablefoottext{3}{the equatorial angular width of the target body full disk, if it were fully
                             visible to the observer}.}
\end{table*}

\section{Discussions}
\label{sec:results}

\begin{table*}
\caption{Summary of all observation-to-model ratios, their scatter, and the number
         of available observations.
\label{tbl:all_ratios}}
\begin{tabular}{llrrrrrrrrrl}
\noalign{\smallskip}\hline\hline\noalign{\smallskip}
       & Obs.\ & \multicolumn{3}{c}{blue band} &  \multicolumn{3}{c}{green band} & \multicolumn{3}{c}{red band} & Model \\
Object & Mode  & average & stdev & num & average & stdev & num   & average & stdev & num  & Version \\
\noalign{\smallskip}\hline\noalign{\smallskip}
Uranus    & CN   & 1.007 & 0.015 &  4 & 1.014 & 0.008 &  4  & 1.041 & 0.002 &  8 & ura\_esa2\_2\_i.dat (Moreno) \\
Uranus    & SM   & 1.011 & 0.005 & 15 & 1.021 & 0.003 & 15  & 1.037 & 0.004 & 25 & ura\_esa2\_2\_i.dat (Moreno) \\
Uranus    & Both & 1.010 & 0.008 & 19 & 1.020 & 0.005 & 19  & 1.038 & 0.004 & 33 & ura\_esa2\_2\_i.dat (Moreno) \\
\noalign{\smallskip}
Uranus    & CN   & 0.917 & 0.013 &  4 & 0.950 & 0.008 &  4  & 0.988 & 0.002 &  8 & orton\_uranus\_esa5 (Orton) \\
Uranus    & SM   & 0.921 & 0.005 & 15 & 0.957 & 0.003 & 15  & 0.984 & 0.004 & 25 & orton\_uranus\_esa5 (Orton) \\
Uranus    & Both & 0.920 & 0.007 & 19 & 0.956 & 0.005 & 19  & 0.985 & 0.004 & 33 & orton\_uranus\_esa5 (Orton) \\
\noalign{\smallskip}
Neptune   & CN   & 0.979 & 0.019 &  5 & 0.988 & 0.016 &  7  & 0.990 & 0.005 & 12 & nep\_esa5\_2\_i.dat (Moreno) \\
Neptune   & SM   & 0.980 & 0.008 & 20 & 0.987 & 0.006 & 28  & 0.990 & 0.006 & 48 & nep\_esa5\_2\_i.dat (Moreno) \\
Neptune   & Both & 0.980 & 0.011 & 25 & 0.987 & 0.009 & 35  & 0.990 & 0.006 & 60 & nep\_esa5\_2\_i.dat (Moreno) \\
\noalign{\smallskip}
Callisto  & SM & 1.042 & 0.003 &  3 &  ---  &  ---  & --- & 1.077 & 0.015 &  3 & call\_esa2\_2\_i.dat (Moreno) \\
Ganymede  & SM & 1.137 & 0.003 &  3 &  ---  &  ---  & --- & 1.210 & 0.030 &  3 & gany\_esa2\_2\_i.dat (Moreno) \\
Titan     & SM & 1.017 & 0.003 &  2 & 0.999 & 0.001 &   2 & 1.010 & 0.004 &  4 & tit\_esa3\_2\_i.dat (Moreno) \\
\noalign{\smallskip}\hline
\end{tabular}
\end{table*}

Table~\ref{tbl:all_ratios} summarizes all observation-to-model ratios, their scatter,
and the number of available observations. Most of the ratios are very close to 1.0,
which shows an excellent agreement between the flux densities derived from the PACS
photometer measurements and the corresponding model predictions. The standard deviations
of these ratios are very small - in many cases well below 1\% - for a given object and 
within a given band. This confirms the very high stability of the PACS bolometer response
over the entire mission lifetime (see also Balog et al.\ \cite{balog14}, M\"uller et al.\ \cite{mueller14},
or Nielbock et al.\ \cite{nielbock13}). The small standard deviations are even more impressive
knowing that the reference measurements on the PACS internal calibration sources were not
even used in the reduction and calibration process (see Mo\'or et al.\ \cite{moor14}).
Neptune observations were taken in low- and high-gain mode (see Tables
\ref{tbl:pacs_chopnod} and \ref{tbl:pacs_scanmap1}),
but no differences in the corresponding observation-to-model ratios can be found.
It is also worth noting that the ratios derived from the chop-nod mode and the scan-map
mode agree within 1\%, showing that the reduction and correction procedures are very well
balanced for these very different observing techniques, and for both gain settings. 

\subsection{Callisto, Ganymede, and Titan}

The Callisto and Ganymede fluxes seem to be systematically higher by 5-20\% than
the corresponding model predictions. In contrast, Titan agrees very well with the 
expected fluxes. We investigated different aspects to determine
whether there are technical reasons
or model shortcomings that might explain the disagreement for Callisto and Ganymede.

The Callisto and Ganymede measurements were taken in non-tracking mode, and to produce
the maps shown in Fig.~\ref{fig:cal_gany}, it was necessary to stack all frames first on
the expected position of Callisto (top right in Fig.~\ref{fig:cal_gany}) and then on
Ganymede (bottom left in Fig.~\ref{fig:cal_gany}). Overall, the final maps look very clean
and flat, and possible stray-light or Jupiter PSF interferences\footnote{Callisto was about
6$^{\prime}$ away from Jupiter, Ganymede only about 5$^{\prime}$.} are entirely
eliminated by the high-pass filtering reduction and the background subtraction
in the aperture photometry step. In addition, for
Titan - with an angular separation of only about 3$^{\prime}$ from Saturn during the PACS measurements -
there are no significant stray-light or PSF contributions in the aperture photometry
(see Fig.~\ref{fig:titan_bgr}). The derived fluxes of the
Saturn and Jupiter satellites are not influenced by the proximity of the planets.

Callisto and Ganymede were separated by 55$^{\prime \prime}$
at the time of the measurements (see Table~\ref{tbl:call_gany_titan}). We calculated the
mutual contribution of the far-field PSF flux distribution within our photometry apertures
of 12$^{\prime \prime}$ and 22$^{\prime \prime}$ in radius. But the flux contribution
of Callisto at the location of Ganymede (and vice versa) is well below 1\% at 70\,$\mu$m
and also at 160\,$\mu$m. The obtained aperture fluxes are therefore reliable despite
the proximity of bright sources.

The measurements were taken with a satellite scan speed of 10$^{\prime \prime}$/s
(instead of the nominal 20$^{\prime \prime}$/s). But the slow scan speed
had no impact on the final fluxes either. We tested the stability of the source
fluxes by using a range of different aperture sizes, but the resulting fluxes were
all within 1\%, confirming the negligible influence of the slower scan speed
(see also D.\ Lutz,
PICC-ME-TN-033\footnote{\tt http://herschel.esac.esa.int/twiki/pub/Public/\-PacsCalibrationWeb/bolopsf\_21.pdf},
Vers.\ 2.1: the baseline for the calculation of aperture encircled energy fractions
was a set of bright-source measurements including observations with 10 and 20$^{\prime \prime}$/s scan speed).

Callisto and Ganymede are at very high flux levels and the non-linearity correction on
signal level is significant. While this correction is in the order of 0-6\% for
asteroids (M\"uller et al.\ \cite{mueller14}) and 5-15\% for the planets Uranus and Neptune,
it is about 17\% for the two Galilean satellites in the blue band and 5-6\% in the red band.
The true uncertainty of this correction is unknown, but it seems to be very reliable
for Uranus and Neptune, for which the non-linearity corrections lead to an excellent 
agreement between observations and models. The non-linearity corrections for Callisto
and Ganymede are only marginally larger, and there is no reason that a
slightly larger correction would introduce additional errors. The non-linearity
corrections are almost identical for Callisto and Ganymede, and therefore they 
cannot account for the offset difference seen between the two targets in Fig.\ref{fig:callisto_ganymede_titan}.
A conservative estimate of the quality of the non-linearity correction leads
to an additional error of $\approx$5\%
when the correction is above 15\% (applicable for Uranus, Neptune, Callisto, and Ganymede
in the blue and green bands) and well below ($\approx$2\%) for smaller corrections
in the red band (see also PICC-NHSC-TR-031\footnote{available from the
Herschel Explanatory Library Listings (HELL) at {\tt http://www.cosmos.esa.int/\-web/\-herschel}}).
These uncertainties - together with the 5\% absolute flux uncertainty and the small uncertainty
introduced by colour correction - have to be considered when using the PACS absolute fluxes
for model adjustments.

The HSA provides additional well-calibrated measurements of Titan and Callisto taken by the
SPIRE spectrometer (Griffin et al.\ \cite{griffin10}; Swinyard et al.\ \cite{swinyard10}).
First, we extracted the Titan spectrometer observations (SPG version 13.0.0) from
OD 404 (OBSID 1342198925) and OD 803 (OBSID 1342224755) and compared the observed
continuum level with the "tit\_esa3\_2\_i.dat" model prediction using the apparent
effective sizes of 0.7460$^{\prime \prime}$ and 0.7131$^{\prime \prime}$, respectively.
On average, the SPIRE spectrometer continuum fluxes of Titan are about 5\%
higher than the corresponding model predictions. This agrees
well with our observation-to-model ratios for Titan (see Tables~\ref{tbl:titan} and \ref{tbl:all_ratios})
considering the 5\% absolute flux accuracy of PACS
and SPIRE and 
the very different calibration schemes for the two instruments.
In a second step, we repeated this procedure for Callisto. The reduced and
calibrated SPIRE spectrometer data from OD 602 (OBSID 1342212340) were extracted
from the HSA. The model calculations based on "call\_esa2\_2\_i.dat" and an
apparent effective size of 1.29$^{\prime \prime}$ led to observation-to-model ratios
of 1.10-1.15, very close to the PACS 160\,$\mu$m ratio of 1.08. This exercise
confirms our confidence in the derived PACS fluxes and indicates
that problems with the models are the main cause of the derived observation-to-model
ratios for Callisto and Ganymede.

A closer inspection of the Callisto and Ganymede model-set\-ups immediately
revealed a first possible cause for the lower model fluxes. The brightness temperatures
in both models refer to a heliocentric distance of 5.21\,AU, but the PACS Callisto 
and Ganymede measurements were taken at 4.98\,AU heliocentric distance. As a consequence,
the temperature at the sub-solar point is higher by a factor of about 1.02, which
influences the model brightness temperatures. This correction would improve the 
agreement between observed Callisto fluxes and model predictions. We estimated that
the new Callisto ratios would then be at 1.05 or below in both PACS bands.
For Ganymede the offset would shrink slightly, but the new observation-to-model
ratios would still be around 1.1 at 70\,$\mu$m and slightly above 1.15 at 160\,$\mu$m.
But Ganymede is in general more difficult to model correctly over the entire
wavelength range (i.e.,\ from IR to radio). Especially the thermal inertias of the
sub-surface layers of Ganymede are not well understood.

To facilitate the usage of our derived and well-calibrated fluxes, we also provide
brightness temperatures (see Col.\ T$_B$ in Tables~\ref{tbl:callisto}, \ref{tbl:ganymede},
and \ref{tbl:titan}) using Planck's law under the assumption of a black body. First,
the apparent effective diameter D is translated into a solid angle
$\Omega$ [sr] = (D [$^{\prime \prime}$] / 2)$^2$ $\cdot$ $\pi$ $\cdot$ 2.3504$\cdot$10$^{-11}$.
This allows calculating the surface brightness
B$_{\nu}$ [MJy/sr] = F$_{\nu}^{obs}$ [Jy] $\cdot$ 1.0$\cdot$10$^{-6}$ $\cdot$ $\Omega$$^{-1}$.
The brightness temperature T$_{b}$ [K] at a given wavelength $\lambda$ [$\mu$m] is then given by\\

T$_{b}^{-1}$ = ln((1$\cdot$10$^{25}$$\cdot$2$\cdot$h$\cdot$c$\cdot$$\lambda$$^{-3}$)/B$_{\nu}$ + 1)$\cdot$ $\frac{k \cdot \lambda}{h \cdot c}$ \\

with the speed of light c = 2.99792458$\cdot$10$^{14}$ $\mu$m/s, the Planck constant h = 6.6262$\cdot$10$^{-27}$\,erg$\cdot$s, and
the Boltzmann constant k = 1.3807$\cdot$10$^{-16}$ erg$\cdot$K$^{-1}$.

The uncertainties in T$_b$ are calculated from the total flux uncertainties:
(i) 5\% in absolute flux calibration; (ii) 1\% or 2\% in colour correction;
(iii) and the non-linearity correction: 0\%, if the correction is well below 5\%,
2\%, if the non-linearity correction is around 5\%, and 5\%, if the correction
is well above 10\%. The summed flux uncertainties are then between 5\% and 7\% for
Callisto, Ganymede, and Titan in the three PACS bands.
Overall, the Titan fluxes agree extremely well with the "tit\_esa3\_2\_i.dat" model by Moreno,
no adjustments are necessary. The Callisto model requires slight modifications to
account for the smaller heliocentric distance of the PACS measurement. We expect that
this will also lead to a good agreement between observations and model within the
given error bars. Only Ganymede shows a mismatch, and the observed fluxes are 10-20\%
higher than model predictions.
Our PACS fluxes are therefore an important legacy and key element for future model
updates, for instance,\ for calibration purposes of ground-based and airborne
sub-millimeter and millimeter projects like ALMA, IRAM, or LMT.

\subsection{Uranus}

The observations of Uranus agree well with the Moreno-based predictions
within the given 5-7\% model accuracy. But the Uranus model of Orton ("orton\_uranus\_esa5")
shows problems in the short-wavelength PACS range. The model predictions are too high by
about 8\% at 70\,$\mu$m and still about 4-5\% too high at 100\,$\mu$m. The overall impression
is that the model seems to work well at the longest PACS wavelength range and also
in the SPIRE range (see Bendo et al.\ \cite{bendo13}, Griffin et al.\ \cite{griffin13}). The reason for the offset 
might be related to the He/H$_2$ ratio used in the model (Orton, priv.\ comm.). Here,
the PACS data will place new and very important constraints on this fundamental property
in the Uranus atmosphere.

\subsection{Neptune}

The observed Neptune fluxes agree within 1-2\% with the Moreno "esa5" model predictions.
Considering that the PACS photometer fluxes are tied to stellar models and the SPIRE
photometer calibration is based on Neptune, our analysis of Neptune shows that
PACS and SPIRE produce reliable fluxes on the same absolute level (well within 5\%).

\subsection{Influence of spectral features}

Several planetary objects show spectral features (notably, Neptune and Titan,
see Fig.~\ref{fig:sed_all}), related to minor species (nitriles and water).
HCN and H$_2$O were taken into account for Neptune's model. H$_2$O has some
strong but narrow lines around 60\,$\mu$m, but with a negligible contribution
of less than 0.01\% of the PACS continuum. The HCN contribution is about ten
times smaller.  
In the model for Titan, CH$_3$CN and HC$_3$N are also narrow and
negligible at wavelengths below 300\,$\mu$m because of their
high rotational J number and lower energy levels $>$ 1000\,cm$^{-1}$
, which are difficult to populate at atmospheric temperatures of about 150\,K.

In summary, these  minor species (nitriles and water) do not contribute
significantly in the PACS range because (i) the linewidths of these
species are very narrow; (ii) the line intensities of CH$_3$CN and HC$_3$N
decrease strongly with decreasing wavelength below 300\,$\mu$m. We
estimated that within the photometer spectral bandwidth, the contribution
is below 0.01\% of the total flux and is therefore negligible.

\subsection{Non-linearity correction}

\begin{figure}[h!tb]
\centering
  \rotatebox{0}{\resizebox{\hsize}{!}{\includegraphics{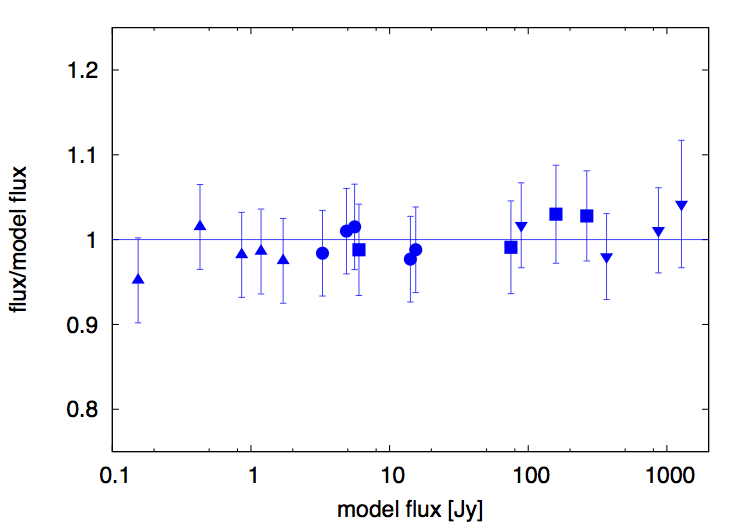}}}
  \caption{Reduced and calibrated blue-band fluxes of faint stars (triangles),
           fiducial stars (circles), prime asteroids (squares), planets (Uranus and Neptune)
           and satellites Callisto and Titan (reversed triangles), divided by their
           corresponding model fluxes and shown as a function of model flux.
\label{fig:nonlin_obsmod_blue}}
\end{figure}

\begin{figure}[h!tb]
\centering
  \rotatebox{0}{\resizebox{\hsize}{!}{\includegraphics{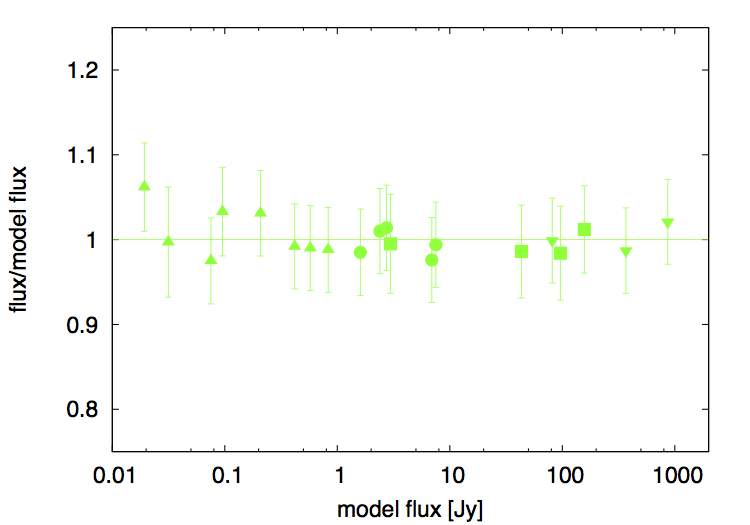}}}
  \caption{Reduced and calibrated green-band fluxes of stars, asteroids, planets, satellites,
           divided by their corresponding model fluxes and shown as a function of model flux.
\label{fig:nonlin_obsmod_green}}
\end{figure}

\begin{figure}[h!tb]
\centering
  \rotatebox{0}{\resizebox{\hsize}{!}{\includegraphics{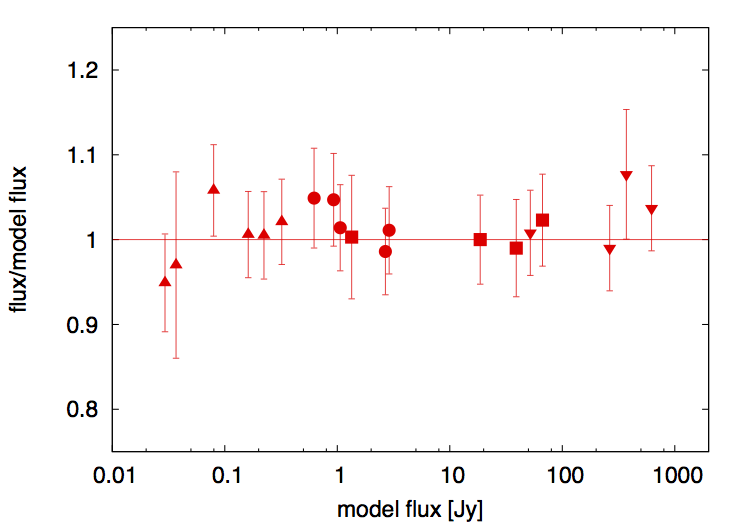}}}
  \caption{Reduced and calibrated red-band fluxes of stars, asteroids, planets, satellites,
           divided by their corresponding model fluxes and shown as a function of model flux.
\label{fig:nonlin_obsmod_red}}
\end{figure}
 
The results for Uranus, Neptune, and Callisto show the reliability of the PACS
fluxes in the high-flux regime where the bolometer response is already in the
non-linear regime. The PACS photometer pipeline products and the
various reduction scripts take care of non-linearity corrections in a reliable
way and point-source fluxes up to 1400\,Jy are validated. The rare cases of even
brighter sources (where non-linearity corrections of larger than 20\% are needed)
are formally not verified. However, the pre-flight pixel-by-pixel non-linearity
response calibration provides at least a good first-order correction, and the derived
flux densities for point-sources above 1400\,Jy up to the ADC\footnote{analog-to-digital converter}
saturation limit are reliable on an estimated 10\% level or better.

In Figs.~\ref{fig:nonlin_obsmod_blue}, \ref{fig:nonlin_obsmod_green}, and
\ref{fig:nonlin_obsmod_red} we show the observation-to-model ratios for objects
spanning 4-5 orders of magnitude in flux. The ratios for the stars are taken
from Balog et al.\ (\cite{balog14}), Nielbock et al.\ (\cite{nielbock13}), and
Klaas et al.\ (in preparation). The ratios for the asteroids are taken from
M\"uller et al.\ (\cite{mueller14}), the remaining from our Table~\ref{tbl:all_ratios}.
In all three PACS bands we find a very good agreement with stellar, asteroid,
planet, and satellite models from well below 100\,mJy to fluxes well above 1000\,Jy.
There are also no offsets between low- and high-gain observations, fixed or
solar system targets, measurements taken in chop-nod or in scan-map mode, or between
PACS and SPIRE photometric measurements.

\subsection{Cross calibration}

From our experience with asteroids (M\"uller et al.\ \cite{mueller14}), and now also
from the work on planets and satellites, we conclude that the stellar-based absolute
flux calibration of the PACS photometer agrees very well with the Neptune-based photometer
calibration of SPIRE (Bendo et al.\ \cite{bendo13}). The PACS photometer data
presented here are crucial for future model updates of Uranus, Callisto,
and Ganymede, which are mainly based on radiometry done by interplanetary
missions. Improved models are needed for high-frequency amplitude calibration of
ALMA and other submm projects. These bright point-sources are also excellent
cross-calibration targets for the PACS and SPIRE spectrometers and for
SOFIA-FORCAST or FIFI-LS as well.

\section{Summary and outlook}
\label{sec:conclusions}

We present the PACS photometer observations of bright satellites and planets,
mainly taken as part of the coordinated instrument calibration program.
These measurements - some of them also taken in non-standard instrument
configurations - show the high reliability of the PACS photometer 
chop-nod and scan-map fluxes: absolute flux densities are given with an absolute flux error of
5\% and a relative (repeatability) error of 1-2\%.
It is important to note here that the absolute
calibration of our derived fluxes is tied to five fiducial stars (Balog et al.\ \cite{balog14}),
and they are therefore completely independent of planet, asteroid, or
satellite model predictions.

The comparison with planet and satellite model predictions shows that the
established corrections - mainly the non-linearity correction, flux correction for chop-nod data,
aperture corrections, or color corrections - are accurate and can be
applied as recommended in the PACS handbook and data reduction guides.
These corrections are also validated up to the highest flux levels just
below the level of ADC saturation.
Overall, the PACS photometer standard reduction and calibration procedures
lead to reliably calibrated flux densities for point sources ranging from a few
milli-Jansky up to about 1000\,Jy.

The observations of Neptune and Titan show excellent agreement - within the 
assigned 5\% absolute accuracy - with available models. The Uranus, Callisto,
and Ganymede observations indicate that the latest available models have some
problems in the PACS wavelength range: The latest Uranus model ("esa5") shows a
wavelength-dependent offset of about 10\% at 70\,$\mu$m, but an acceptable
agreement (within 5\%) at 160\,$\mu$m. For Callisto it is the opposite: the
model shows a 5\% agreement at 70\,$\mu$m and is about 10\% too low at
the long wavelengths. The Ganymede model underestimates the observed and
calibrated fluxes by 15 to 25\%.

The final tabulated fluxes and the observed brightness temperatures can now
be used for future model updates and dedicated studies of surface and atmosphere
effects of planets and satellites. Improved model solutions are also relevant
in the context of cross calibration between space-based, airborne, and ground-based facilities.
These bright point sources are also very important for various calibration
activities for ALMA and other astronomical sub-millimeter and millimeter projects.

\begin{acknowledgements}
The entire PACS-ICC calibration team is acknowledged for countless
contributions in the preparation, conduction, and evaluation process
related to the observations of bright point sources.
We would also like to thank Nicolas Billot for his feedback
on the non-linearity corrections, which are based on 
PACS pre-launch test campaigns in 2006-2007, and small
adjustments for the final in-flight detector bias settings. Glenn Orton,
Martin Burgdorf, and Anthony Marston provided help with the planet and
satellite models and contributed to the discussions of the results.
A.\ Mo\'or was supported by the Momentum grant of the MTA CSFK
Lend\"ulet Disk Research Group.
PACS has been developed by a consortium of institutes led by MPE (Germany)
and including UVIE (Austria); KUL, CSL, IMEC (Belgium); CEA, OAMP (France);
MPIA (Germany); IFSI, OAP/AOT, OAA/CAISMI, LENS, SISSA (Italy); IAC (Spain).
This development has been supported by the funding agencies BMVIT (Austria),
ESA-PRODEX (Belgium), CEA/CNES (France), DLR (Germany), ASI (Italy),
and CICYT/MCYT (Spain). SPIRE has been developed by a consortium of institutes
led by Cardiff University (UK) and including Univ.\ Lethbridge (Canada);
NAOC (China); CEA, LAM (France); IFSI, Univ.\ Padua (Italy); IAC (Spain); 
Stockholm Observatory (Sweden); Imperial College London, RAL, UCL-MSSL, UKATC,
Univ.\ Sussex (UK); and Caltech, JPL, NHSC, Univ. Colorado (USA). This
development has been supported by national funding agencies: CSA (Canada);
NAOC (China); CEA, CNES, CNRS (France); ASI (Italy); MCINN (Spain);
SNSB (Sweden); STFC, UKSA (UK); and NASA (USA).
This work was supported by the Hungarian OTKA grant K104607.
We thank the anonymous referee for constructive comments.
\end{acknowledgements}

\begin{figure*}[h!tb]
\centering
  \rotatebox{0}{\resizebox{\hsize}{!}{\includegraphics{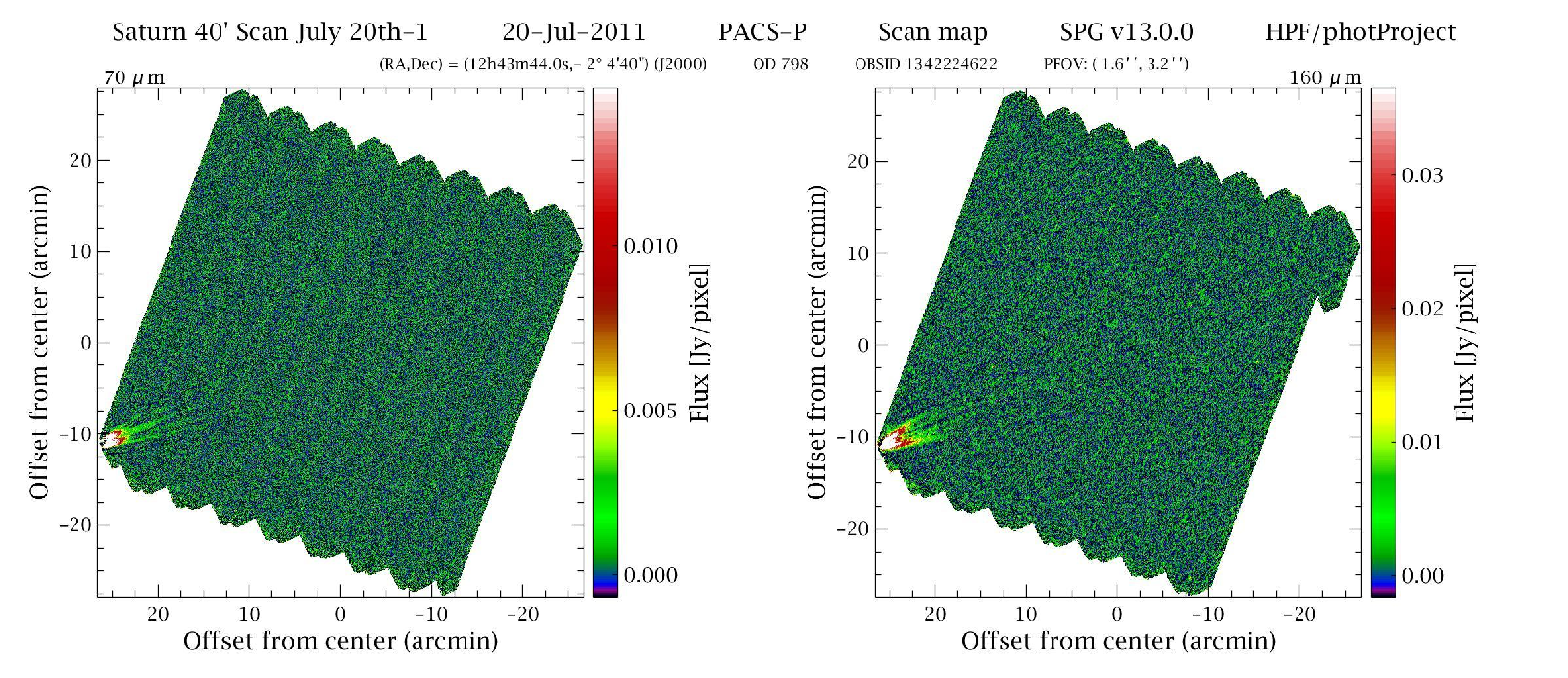}}}
  \rotatebox{0}{\resizebox{\hsize}{!}{\includegraphics{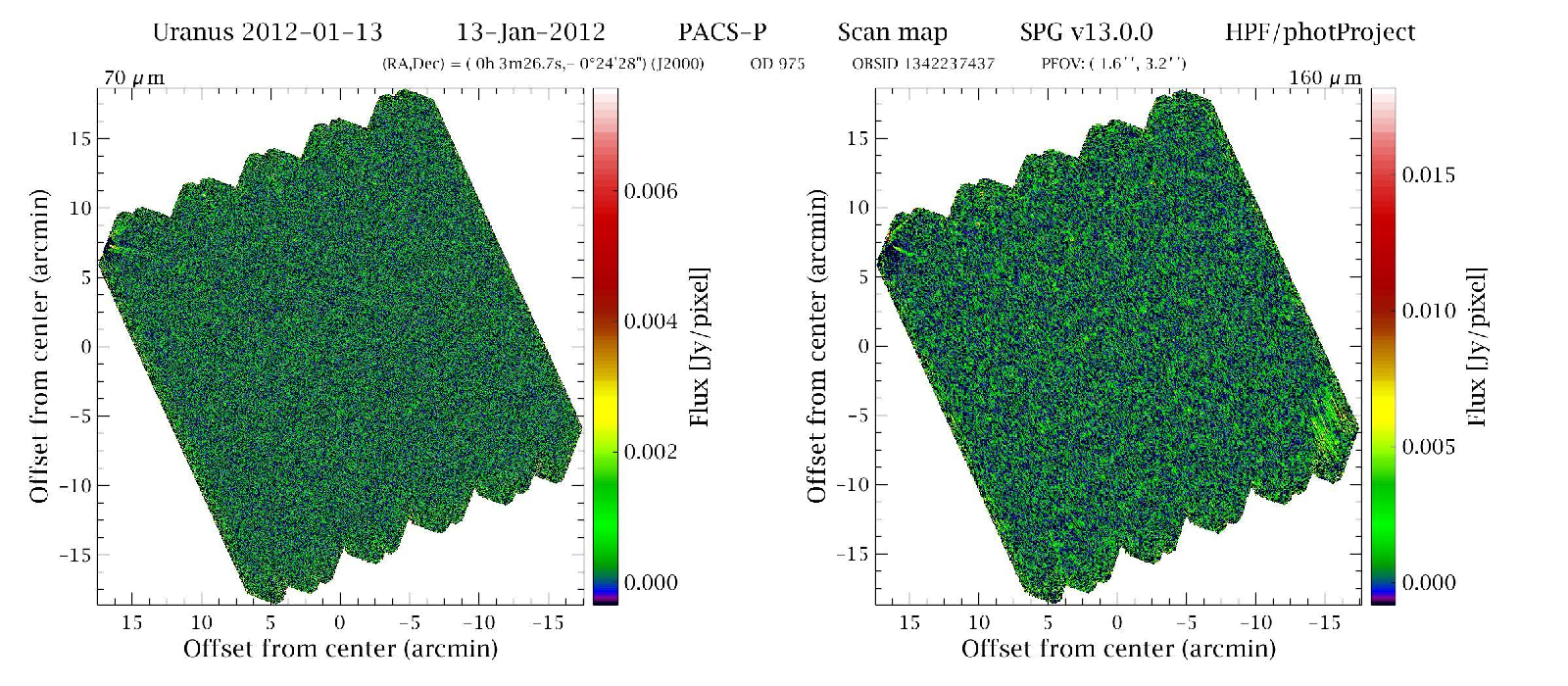}}}
  \caption{Top: HSA maps of OBSID 1342224622 with Saturn located outside the field of view in the lower left corner.
           Bottom: HSA maps of OBSID 1342237437 with Uranus located outside the field of view beyond the upper left corner.
           These measurements are only useful for off-source investigations in the planet's direct vicinity.
\label{fig:outside}}
\end{figure*}

\clearpage
\newpage
\begin{appendix}
\begin{onecolumn}

\section{Bright-source measurements}
\label{sec:obsapp}

In the following tables we list the available photometric observations
of the bright sources, mainly related to calibration programs. Only
Uranus and Neptune were observed in the PACS photometer chop-nod observing
mode (point-source mode) with dithering (Table~\ref{tbl:pacs_chopnod}).
Both sources were measured in all three bands, some in low, some in
high gain, each time only in a single repetition of the pre-defined
dithered chop-nod pattern.

On-source scan-map observations were taken for Mars, Uranus, Neptune,
Callisto, Ganymede, and Titan (Table~\ref{tbl:pacs_scanmap1}). These measurements
cover different instrument settings (bands, gain, scan-speed, map settings),
and different fields on sky. The source is usually close to the map center
(solar system object tracking mode), only one set of Neptune measurements was
taken in fixed mode with the source located off-center, but still well
within the final maps. The Callisto and Ganymede observations were also taken
in fixed mode, but with both sources close to the map center.
The Mars measurements are heavily saturated, and it
is not possible to extract meaningful fluxes.

In the HSA one set of measurements is labeled "Saturn scan" and another "Uranus 2012-01-07" and "Uranus 2012-01-13". They were taken in fixed mode and
as part of a science project (Table~\ref{tbl:pacs_scanmap2}). In these measurements
only off-source fields were scanned and only parts of the PSF wings are visible in
one corner of the maps (Figure~\ref{fig:outside}). The data are part of a 
science program aiming at the detection of dust rings from irregular satellites.

\end{onecolumn}
\clearpage

\begin{sidewaystable*}
\caption{Herschel-PACS photometer {\bf chop-nod} observations (proposals Calibration\_pvpacs\_\#\# and Calibration\_rppacs\_\#\#\#),
        taken in point-source observing mode with dithering and solar system object (SSO) tracking mode.
        SAA: solar aspect angle; dur.: duration of observation in seconds; fil.: filter/band combination               
        (B: 70/160\,$\mu$m; G: 100/160\,$\mu$m); G.: low (L) or high (H) gain; R: repetition of entire chop-nod pattern;
        r: heliocentric distance; $\Delta$: Herschel-centric distance; $\alpha$: phase angle; diam.: effective angular diameter\tablefootmark{1}.
\label{tbl:pacs_chopnod}}
\begin{tabular}{rllrllrcrlrrrr}
\noalign{\smallskip}\hline\hline\noalign{\smallskip}
   &       &        & SAA          & \multicolumn{2}{c}{UTC Start time} & dur. &      &    &    & r    & $\Delta$ & $\alpha$     & diam.\tablefootmark{1} \\
OD & OBSID & Target & [$^{\circ}$] & yyyy mon dd     &  hh:mm:ss        & [s]  & fil. & G. & R. & [AU] & [AU]     & [$^{\circ}$] & [$^{\prime \prime}$] \\
\noalign{\smallskip}\hline\noalign{\smallskip}
 212 & 1342188056 &  Uranus &   -1.9 & 2009 Dec 12 & 14:52:22 & 171 & B/R & L & 1 & 20.09716 & 20.04905 & 3.00 & 3.476 \\ 
 212 & 1342188057 &  Uranus &   -1.9 & 2009 Dec 12 & 14:56:16 & 171 & G/R & L & 1 & 20.09716 & 20.04909 & 3.00 & 3.476 \\ 
 579 & 1342211116 &  Uranus &   -5.6 & 2010 Dec 13 & 17:09:40 & 162 & B/R & L & 1 & 20.08994 & 19.97489 & 2.83 & 3.487 \\ 
 579 & 1342211119 &  Uranus &   -5.0 & 2010 Dec 13 & 17:26:36 & 162 & G/R & L & 1 & 20.08994 & 19.97509 & 2.83 & 3.487 \\ 
 789 & 1342223981 &  Uranus &  -14.3 & 2011 Jul 12 & 01:15:35 & 162 & B/R & L & 1 & 20.08320 & 19.80140 & 2.84 & 3.524 \\ 
 789 & 1342223984 &  Uranus &  -14.4 & 2011 Jul 12 & 01:32:07 & 162 & G/R & L & 1 & 20.08320 & 19.80121 & 2.84 & 3.524 \\ 
 957 & 1342235628 &  Uranus &   -5.1 & 2011 Dec 26 & 22:27:18 & 162 & B/R & L & 1 & 20.07648 & 20.11786 & 2.83 & 3.464 \\ 
 957 & 1342235631 &  Uranus &    4.0 & 2011 Dec 26 & 22:55:25 & 162 & G/R & L & 1 & 20.07648 & 20.11820 & 2.83 & 3.464 \\ 
\noalign{\smallskip}
\noalign{\smallskip}
 173 & 1342186637 & Neptune  & -10.5 & 2009 Nov 03 & 01:35:52 & 171 & B/R & H & 1 & 30.02612 & 29.79070 & 1.86 & 2.277 \\ 
 173 & 1342186638 & Neptune  & -12.5 & 2009 Nov 03 & 01:45:23 & 171 & G/R & H & 1 & 30.02611 & 29.79081 & 1.86 & 2.277 \\ 
 173 & 1342186643 & Neptune  & -12.5 & 2009 Nov 03 & 02:05:25 & 171 & G/R & L & 1 & 30.02611 & 29.79104 & 1.86 & 2.277 \\ 
 212 & 1342188052 & Neptune  &  26.7 & 2009 Dec 12 & 14:26:45 & 171 & G/R & H & 1 & 30.02504 & 30.45664 & 1.70 & 2.227 \\ 
 212 & 1342188055 & Neptune  &  26.7 & 2009 Dec 12 & 14:43:15 & 171 & B/R & H & 1 & 30.02504 & 30.45682 & 1.70 & 2.227 \\ 
 371 & 1342196724 & Neptune  &  -0.1 & 2010 May 19 & 23:19:09 & 162 & G/R & H & 1 & 30.02059 & 29.99826 & 1.95 & 2.262 \\ 
 540 & 1342209039 & Neptune  & -13.0 & 2010 Nov 05 & 01:02:11 & 162 & G/R & H & 1 & 30.01563 & 29.77119 & 1.86 & 2.278 \\ 
 540 & 1342209042 & Neptune  & -13.0 & 2010 Nov 05 & 01:17:13 & 162 & B/R & L & 1 & 30.01563 & 29.77137 & 1.86 & 2.278 \\ 
 716 & 1342220894 & Neptune  &  21.3 & 2011 Apr 29 & 23:36:19 & 162 & B/R & L & 1 & 30.01024 & 30.36190 & 1.81 & 2.235 \\ 
 716 & 1342220897 & Neptune  &  21.3 & 2011 Apr 29 & 23:52:12 & 162 & G/R & H & 1 & 30.01024 & 30.36173 & 1.81 & 2.235 \\ 
 919 & 1342232522 & Neptune  &  -1.5 & 2011 Nov 19 & 03:21:00 & 162 & B/R & L & 1 & 30.00382 & 29.95799 & 1.91 & 2.264 \\ 
 919 & 1342232525 & Neptune  &  -1.5 & 2011 Nov 19 & 03:36:54 & 162 & G/R & H & 1 & 30.00381 & 29.95818 & 1.91 & 2.264 \\ 
\noalign{\smallskip}\hline
\end{tabular}
\tablefoot{\tablefoottext{1}{Taken from "diam.dat", provided by R.\ Moreno, Jul.\ 25, 2012 to the Herschel Calibration Steering Group Team,
            taking into account the equatorial and polar radii and the sub-observer latitude.}}
\end{sidewaystable*}

\begin{landscape}
\begin{longtable}{rllrllrclrccrrrrl}
\caption[]{Herschel-PACS photometer {\bf scan-map} observations (proposal Calibration\_pvpacs\_\#\#, Calibration\_rppacs\_\#\#\#,
        OT1\_ddan01\_1), taken in large-scan observing mode, SSO tracking (exceptions see footnote), i.e.,\ with the target
        located well within the observed field.
        SAA: solar aspect angle; dur.: duration of observation in seconds;
        fil.: filter/band combination (B: 70/160\,$\mu$m; G: 100/160\,$\mu$m);
        G.: low (L) or high (H) gain; R.: repetition of entire scan map;
        S.: scan-speed in $^{\prime \prime}$/s,
        scan-map parameters: Len: scan-leg length (in arc\,min)
        $\times$ n: number of scan legs $\times$ sep: scan-leg separation (in arc\,sec);
        ang.: satellite scan angle in degrees with respect to instrument reference frame;
        r: heliocentric distance; $\Delta$: Herschel-centric distance; $\alpha$: phase angle;
        diam: effective angular diameter\tablefootmark{1}.} \\
\hline\noalign{\smallskip}
   &       &        & SAA          & \multicolumn{2}{c}{UTC Start time} & dur. &      &    &    & S.                     & Len$\times$n$\times$sep                         & ang.         & r    & $\Delta$ & $\alpha$     & diam\tablefootmark{1} \\
OD & OBSID & Target & [$^{\circ}$] & yyyy mon dd & hh:mm:ss             & [s]  & fil. & G. & R. & [$^{\prime \prime}$/s] & [$^{\prime}$ $\times$ \# $\times$ $^{\prime \prime}$] & [$^{\circ}$] & [AU] & [AU]     & [$^{\circ}$] & [$^{\prime \prime}$]  \\
\noalign{\smallskip}\hline\noalign{\smallskip} \endfirsthead                        
\caption[]{\emph{continued}} \\                                                     
\hline\noalign{\smallskip}                                                                                  
   &       &        & SAA          & \multicolumn{2}{c}{UTC Start time} & dur. &      &    &    & S.                     & Len$\times$n$\times$sep                         & ang.         & r    & $\Delta$ & $\alpha$     & diam\tablefootmark{1} \\
OD & OBSID & Target & [$^{\circ}$] & yyyy mon dd & hh:mm:ss             & [s]  & fil. & G. & R. & [$^{\prime \prime}$/s] & [$^{\prime}$ $\times$ \# $\times$ $^{\prime \prime}$] & [$^{\circ}$] & [AU] & [AU]     & [$^{\circ}$] & [$^{\prime \prime}$]  \\
\noalign{\smallskip}\hline\noalign{\smallskip}  \endhead                                                    
\hline                                                                                                      
\multicolumn{17}{c}{\emph{continued on next page}} \endfoot                                                 
\noalign{\smallskip}\hline \endlastfoot                                                                     
\label{tbl:pacs_scanmap1}
 888 & 1342231157\tablefootmark{2} & Mars  &   22.8 & 2011 Oct 19 & 06:44:59 & 3729 & B/R & H & 1 & 60 & 80 $\times$ 21 $\times$ 20  &  45 & 1.59877 & 1.68205 & 35.60 & 5.568 \\  
 888 & 1342231158\tablefootmark{2} & Mars  &   22.1 & 2011 Oct 19 & 07:49:26 & 3729 & B/R & H & 1 & 60 & 80 $\times$ 21 $\times$ 20  & 135 & 1.59882 & 1.68171 & 35.61 & 5.569 \\  
 888 & 1342231159\tablefootmark{2} & Mars  &   22.1 & 2011 Oct 19 & 08:52:45 & 3729 & B/R & H & 1 & 60 & 80 $\times$ 21 $\times$ 20  & 225 & 1.59886 & 1.68138 & 35.61 & 5.570 \\  
 888 & 1342231160\tablefootmark{2} & Mars  &   22.8 & 2011 Oct 19 & 09:56:04 & 3729 & B/R & H & 1 & 60 & 80 $\times$ 21 $\times$ 20  & 315 & 1.59891 & 1.68104 & 35.61 & 5.571 \\  
 888 & 1342231161\tablefootmark{2} & Mars  &   22.8 & 2011 Oct 19 & 10:59:23 & 3729 & G/R & H & 1 & 60 & 80 $\times$ 21 $\times$ 20  &  45 & 1.59895 & 1.68071 & 35.62 & 5.572 \\  
 888 & 1342231162\tablefootmark{2} & Mars  &   22.0 & 2011 Oct 19 & 12:02:42 & 3729 & G/R & H & 1 & 60 & 80 $\times$ 21 $\times$ 20  & 135 & 1.59900 & 1.68037 & 35.62 & 5.573 \\  
 888 & 1342231163\tablefootmark{2} & Mars  &   22.1 & 2011 Oct 19 & 13:06:01 & 3729 & G/R & H & 1 & 60 & 80 $\times$ 21 $\times$ 20  & 225 & 1.59905 & 1.68004 & 35.63 & 5.574 \\  
 888 & 1342231164\tablefootmark{2} & Mars  &   22.7 & 2011 Oct 19 & 14:09:21 & 3729 & G/R & H & 1 & 60 & 80 $\times$ 21 $\times$ 20  & 315 & 1.59909 & 1.67971 & 35.63 & 5.576 \\  
 906 & 1342231949\tablefootmark{2} & Mars  &   14.5 & 2011 Nov 06 & 04:07:04 & 3245 & B/R & H & 1 & 20 & 20 $\times$ 41 $\times$  6  &  90 & 1.61655 & 1.53977 & 36.93 & 6.082 \\  
 906 & 1342231950\tablefootmark{2} & Mars  &   14.5 & 2011 Nov 06 & 05:05:20 & 3245 & G/R & H & 1 & 20 & 20 $\times$ 41 $\times$  6  &  90 & 1.61658 & 1.53944 & 36.93 & 6.084 \\  
\noalign{\smallskip}
 579 & 1342211117 & Uranus   &   -5.2 & 2010 Dec 13 & 17:15:17 & 286  & B/R & L  & 1 & 20 & 3.0 $\times$ 10 $\times$  4 &  70 & 20.08994 & 19.97497 & 2.83 & 3.487 \\  
 579 & 1342211118 & Uranus   &   -5.2 & 2010 Dec 13 & 17:21:06 & 286  & B/R & L  & 1 & 20 & 3.0 $\times$ 10 $\times$  4 & 110 & 20.08994 & 19.97504 & 2.83 & 3.487 \\  
 579 & 1342211120 & Uranus   &   -5.2 & 2010 Dec 13 & 17:30:40 & 286  & G/R & L  & 1 & 20 & 3.0 $\times$ 10 $\times$  4 &  70 & 20.08994 & 19.97515 & 2.83 & 3.487 \\  
 579 & 1342211121 & Uranus   &   -5.2 & 2010 Dec 13 & 17:36:29 & 286  & G/R & L  & 1 & 20 & 3.0 $\times$ 10 $\times$  4 & 110 & 20.08994 & 19.97522 & 2.83 & 3.487 \\  
 789 & 1342223982 & Uranus   &  -14.5 & 2011 Jul 12 & 01:21:03 & 286  & B/R & L  & 1 & 20 & 3.0 $\times$ 10 $\times$  4 &  70 & 20.08320 & 19.80132 & 2.84 & 3.524 \\  
 789 & 1342223983 & Uranus   &  -14.5 & 2011 Jul 12 & 01:26:52 & 286  & B/R & L  & 1 & 20 & 3.0 $\times$ 10 $\times$  4 & 110 & 20.08320 & 19.80126 & 2.84 & 3.524 \\  
 789 & 1342223985 & Uranus   &  -14.5 & 2011 Jul 12 & 01:36:26 & 286  & G/R & L  & 1 & 20 & 3.0 $\times$ 10 $\times$  4 &  70 & 20.08320 & 19.80115 & 2.84 & 3.524 \\  
 789 & 1342223986 & Uranus   &  -14.5 & 2011 Jul 12 & 01:42:15 & 286  & G/R & L  & 1 & 20 & 3.0 $\times$ 10 $\times$  4 & 110 & 20.08320 & 19.80108 & 2.84 & 3.524 \\  
 957 & 1342235629 & Uranus   &    3.8 & 2011 Dec 26 & 22:44:31 & 286  & B/R & L  & 1 & 20 & 3.0 $\times$ 10 $\times$  4 &  70 & 20.07648 & 20.11808 & 2.83 & 3.464 \\  
 957 & 1342235630 & Uranus   &    3.8 & 2011 Dec 26 & 22:50:20 & 286  & B/R & L  & 1 & 20 & 3.0 $\times$ 10 $\times$  4 & 110 & 20.07648 & 20.11815 & 2.83 & 3.464 \\  
 957 & 1342235632 & Uranus   &    3.8 & 2011 Dec 26 & 22:59:54 & 286  & G/R & L  & 1 & 20 & 3.0 $\times$ 10 $\times$  4 &  70 & 20.07648 & 20.11826 & 2.83 & 3.464 \\  
 957 & 1342235633 & Uranus   &    3.8 & 2011 Dec 26 & 23:05:43 & 286  & G/R & L  & 1 & 20 & 3.0 $\times$ 10 $\times$  4 & 110 & 20.07648 & 20.11833 & 2.83 & 3.464 \\  
1121 & 1342246772 & Uranus   &   20.5 & 2012 Jun 08 & 04:24:01 & 286  & B/R & L  & 1 & 20 & 3.0 $\times$ 10 $\times$  4 &  70 & 20.06880 & 20.40446 & 2.74 & 3.421 \\  
1121 & 1342246773 & Uranus   &   20.5 & 2012 Jun 08 & 04:29:50 & 286  & B/R & L  & 1 & 20 & 3.0 $\times$ 10 $\times$  4 & 110 & 20.06880 & 20.40439 & 2.74 & 3.421 \\  
1121 & 1342246774 & Uranus   &   20.5 & 2012 Jun 08 & 04:35:39 & 286  & G/R & L  & 1 & 20 & 3.0 $\times$ 10 $\times$  4 &  70 & 20.06880 & 20.40433 & 2.74 & 3.421 \\  
1121 & 1342246775 & Uranus   &   20.5 & 2012 Jun 08 & 04:41:28 & 286  & G/R & L  & 1 & 20 & 3.0 $\times$ 10 $\times$  4 & 110 & 20.06880 & 20.40427 & 2.74 & 3.421 \\  
1310 & 1342257193 & Uranus   &  -12.3 & 2012 Dec 14 & 01:43:37 & 286  & B/R & L  & 1 & 20 & 3.0 $\times$ 10 $\times$  4 &  70 & 20.05868 & 19.82251 & 2.77 & 3.517 \\  
1310 & 1342257194 & Uranus   &  -12.3 & 2012 Dec 14 & 01:49:26 & 286  & B/R & L  & 1 & 20 & 3.0 $\times$ 10 $\times$  4 & 110 & 20.05868 & 19.82258 & 2.77 & 3.517 \\  
1310 & 1342257195 & Uranus   &  -12.3 & 2012 Dec 14 & 01:55:15 & 286  & G/R & L  & 1 & 20 & 3.0 $\times$ 10 $\times$  4 &  70 & 20.05868 & 19.82265 & 2.77 & 3.517 \\  
1310 & 1342257196 & Uranus   &  -12.3 & 2012 Dec 14 & 02:01:04 & 286  & G/R & L  & 1 & 20 & 3.0 $\times$ 10 $\times$  4 & 110 & 20.05868 & 19.82272 & 2.77 & 3.517 \\  
\noalign{\smallskip}
 173 & 1342186639 & Neptune  &  -12.5 & 2009 Nov 03 & 01:49:21 &  179 & B/R & L & 1 & 20 & 5.0 $\times$  4 $\times$ 51 &  45 & 30.02611 & 29.79086 & 1.86 & 2.277 \\  
 173 & 1342186640 & Neptune  &  -12.5 & 2009 Nov 03 & 01:53:23 &  179 & B/R & L & 1 & 20 & 5.0 $\times$  4 $\times$ 51 & 135 & 30.02611 & 29.79090 & 1.86 & 2.277 \\  
 173 & 1342186641 & Neptune  &  -12.4 & 2009 Nov 03 & 01:57:25 &  179 & G/R & H & 1 & 20 & 5.0 $\times$  4 $\times$ 51 &  45 & 30.02611 & 29.79095 & 1.86 & 2.277 \\  
 173 & 1342186642 & Neptune  &  -12.5 & 2009 Nov 03 & 02:01:27 &  179 & G/R & H & 1 & 20 & 5.0 $\times$  4 $\times$ 51 & 135 & 30.02611 & 29.79100 & 1.86 & 2.277 \\  
 212 & 1342188050 & Neptune  &   26.7 & 2009 Dec 12 & 14:15:21 &  315 & G/R & H & 1 & 20 & 4.0 $\times$  8 $\times$  4 &  63 & 30.02504 & 30.45653 & 1.70 & 2.227 \\  
 212 & 1342188051 & Neptune  &   26.7 & 2009 Dec 12 & 14:21:39 &  315 & G/R & H & 1 & 20 & 4.0 $\times$  8 $\times$  4 & 117 & 30.02504 & 30.45660 & 1.70 & 2.227 \\  
 212 & 1342188053 & Neptune  &   26.7 & 2009 Dec 12 & 14:31:51 &  315 & B/R & H & 1 & 20 & 4.0 $\times$  8 $\times$  4 &  63 & 30.02504 & 30.45671 & 1.70 & 2.227 \\  
 212 & 1342188054 & Neptune  &   26.7 & 2009 Dec 12 & 14:38:09 &  315 & B/R & H & 1 & 20 & 4.0 $\times$  8 $\times$  4 & 117 & 30.02504 & 30.45678 & 1.70 & 2.227 \\  
 371 & 1342196725 & Neptune  &   -0.1 & 2010 May 19 & 23:23:50 &  276 & G/R & H & 1 & 20 & 2.5 $\times$ 10 $\times$  4 &  70 & 30.02059 & 29.99819 & 1.95 & 2.262 \\  
 371 & 1342196726 & Neptune  &   -0.1 & 2010 May 19 & 23:29:29 &  276 & G/R & H & 1 & 20 & 2.5 $\times$ 10 $\times$  4 & 110 & 30.02059 & 29.99813 & 1.95 & 2.262 \\  
 371 & 1342196727 & Neptune  &   -0.1 & 2010 May 19 & 23:35:08 &  276 & B/R & L & 1 & 20 & 2.5 $\times$ 10 $\times$  4 &  70 & 30.02059 & 29.99806 & 1.95 & 2.262 \\  
 371 & 1342196728 & Neptune  &   -0.1 & 2010 May 19 & 23:40:47 &  276 & B/R & L & 1 & 20 & 2.5 $\times$ 10 $\times$  4 & 110 & 30.02059 & 29.99799 & 1.95 & 2.262 \\  
 540 & 1342209040 & Neptune  &  -13.0 & 2010 Nov 05 & 01:06:52 &  276 & G/R & H & 1 & 20 & 2.5 $\times$ 10 $\times$  4 &  70 & 30.01563 & 29.77126 & 1.86 & 2.278 \\  
 540 & 1342209041 & Neptune  &  -13.0 & 2010 Nov 05 & 01:12:31 &  276 & G/R & H & 1 & 20 & 2.5 $\times$ 10 $\times$  4 & 110 & 30.01563 & 29.77133 & 1.86 & 2.278 \\  
 540 & 1342209043 & Neptune  &  -13.0 & 2010 Nov 05 & 01:21:55 &  276 & B/R & L & 1 & 20 & 2.5 $\times$ 10 $\times$  4 &  70 & 30.01563 & 29.77144 & 1.86 & 2.278 \\  
 540 & 1342209044 & Neptune  &  -13.0 & 2010 Nov 05 & 01:27:34 &  276 & B/R & L & 1 & 20 & 2.5 $\times$ 10 $\times$  4 & 110 & 30.01563 & 29.77150 & 1.86 & 2.278 \\  
 716 & 1342220895 & Neptune  &   21.3 & 2011 Apr 29 & 23:41:36 &  286 & B/R & L & 1 & 20 & 3.0 $\times$ 10 $\times$  4 &  70 & 30.01024 & 30.36183 & 1.81 & 2.235 \\  
 716 & 1342220896 & Neptune  &   21.3 & 2011 Apr 29 & 23:47:25 &  286 & B/R & L & 1 & 20 & 3.0 $\times$ 10 $\times$  4 & 110 & 30.01024 & 30.36177 & 1.81 & 2.235 \\  
 716 & 1342220898 & Neptune  &   21.3 & 2011 Apr 29 & 23:56:58 &  286 & G/R & H & 1 & 20 & 3.0 $\times$ 10 $\times$  4 &  70 & 30.01024 & 30.36167 & 1.81 & 2.235 \\  
 716 & 1342220899 & Neptune  &   21.3 & 2011 Apr 30 & 00:02:47 &  286 & G/R & H & 1 & 20 & 3.0 $\times$ 10 $\times$  4 & 110 & 30.01024 & 30.36160 & 1.81 & 2.235 \\  
 739 & 1342221604 & Neptune  &   -0.0 & 2011 May 23 & 13:24:14 &  286 & B/R & L & 1 & 20 & 3.0 $\times$ 10 $\times$  4 &  70 & 30.00950 & 29.97101 & 1.96 & 2.264 \\  
 739 & 1342221605 & Neptune  &   -1.0 & 2011 May 23 & 13:34:17 &  286 & B/R & L & 1 & 20 & 3.0 $\times$ 10 $\times$  4 & 110 & 30.00950 & 29.97089 & 1.96 & 2.264 \\  
 739 & 1342221606 & Neptune  &   -1.0 & 2011 May 23 & 13:40:05 &  286 & G/R & H & 1 & 20 & 3.0 $\times$ 10 $\times$  4 &  70 & 30.00950 & 29.97082 & 1.96 & 2.264 \\  
 739 & 1342221607 & Neptune  &   -1.0 & 2011 May 23 & 13:45:48 &  286 & G/R & H & 1 & 20 & 3.0 $\times$ 10 $\times$  4 & 110 & 30.00950 & 29.97075 & 1.96 & 2.264 \\  
 759 & 1342222561 & Neptune  &  -19.6 & 2011 Jun 12 & 04:06:37 & 2996 & G/R & H & 4 & 20 & 7.0 $\times$ 18 $\times$ 25 &  45 & 30.00889 & 29.64553 & 1.85 & 2.289 \\  
 759 & 1342222562 & Neptune  &  -19.7 & 2011 Jun 12 & 04:57:46 & 2996 & G/R & H & 4 & 20 & 7.0 $\times$ 18 $\times$ 25 & 135 & 30.00889 & 29.64497 & 1.85 & 2.289 \\  
 759 & 1342222563 & Neptune  &  -19.7 & 2011 Jun 12 & 05:48:45 & 2996 & G/R & H & 4 & 20 & 7.0 $\times$ 18 $\times$ 25 &  45 & 30.00889 & 29.64440 & 1.85 & 2.289 \\  
 759 & 1342222564 & Neptune  &  -19.8 & 2011 Jun 12 & 06:39:44 & 2996 & G/R & H & 4 & 20 & 7.0 $\times$ 18 $\times$ 25 & 135 & 30.00888 & 29.64384 & 1.85 & 2.289 \\  
 919 & 1342232523 & Neptune  &   -1.5 & 2011 Nov 19 & 03:26:17 &  286 & B/R & L & 1 & 20 & 3.0 $\times$ 10 $\times$  4 &  70 & 30.00381 & 29.95807 & 1.91 & 2.264 \\  
 919 & 1342232524 & Neptune  &   -1.5 & 2011 Nov 19 & 03:32:06 &  286 & B/R & L & 1 & 20 & 3.0 $\times$ 10 $\times$  4 & 110 & 30.00381 & 29.95814 & 1.91 & 2.264 \\  
 919 & 1342232526 & Neptune  &   -1.5 & 2011 Nov 19 & 03:41:39 &  286 & G/R & H & 1 & 20 & 3.0 $\times$ 10 $\times$  4 &  70 & 30.00381 & 29.95825 & 1.91 & 2.264 \\  
 919 & 1342232527 & Neptune  &   -1.5 & 2011 Nov 19 & 03:47:28 &  286 & G/R & H & 1 & 20 & 3.0 $\times$ 10 $\times$  4 & 110 & 30.00381 & 29.95832 & 1.91 & 2.264 \\  
 936 & 1342234207\tablefootmark{3} & Neptune & 14.8 & 2011 Dec 05 & 13:00:23 & 2986 & G/R & H & 4 & 20 & 7.0 $\times$ 18 $\times$ 25 &  45 & 30.00329 & 30.23924 & 1.84 & 2.243 \\ 
 936 & 1342234208\tablefootmark{3} & Neptune & 14.9 & 2011 Dec 05 & 13:54:50 & 2986 & G/R & H & 4 & 20 & 7.0 $\times$ 18 $\times$ 25 & 135 & 30.00329 & 30.23988 & 1.84 & 2.243 \\ 
 947 & 1342234435\tablefootmark{3} & Neptune & 26.0 & 2011 Dec 16 & 19:47:29 & 2986 & G/R & H & 4 & 20 & 7.0 $\times$ 18 $\times$ 25 &  45 & 30.00293 & 30.42232 & 1.71 & 2.229 \\ 
 947 & 1342234436\tablefootmark{3} & Neptune & 26.0 & 2011 Dec 16 & 20:40:39 & 2986 & G/R & H & 4 & 20 & 7.0 $\times$ 18 $\times$ 25 & 135 & 30.00293 & 30.42290 & 1.71 & 2.229 \\ 
1097 & 1342245787\tablefootmark{4} & Neptune &  8.2 & 2012 May 15 & 07:25:59 &  286 & G/R & H & 1 & 20 & 3.0 $\times$ 10 $\times$  4 &  70 & 29.99809 & 30.12620 & 1.93 & 2.252 \\ 
1097 & 1342245788\tablefootmark{4} & Neptune &  8.1 & 2012 May 15 & 07:31:48 &  286 & G/R & H & 1 & 20 & 3.0 $\times$ 10 $\times$  4 & 110 & 29.99809 & 30.12614 & 1.93 & 2.252 \\ 
1097 & 1342245789\tablefootmark{4} & Neptune &  8.1 & 2012 May 15 & 07:37:37 &  286 & B/R & L & 1 & 20 & 3.0 $\times$ 10 $\times$  4 &  70 & 29.99809 & 30.12607 & 1.93 & 2.252 \\ 
1097 & 1342245790\tablefootmark{4} & Neptune &  8.1 & 2012 May 15 & 07:43:26 &  286 & B/R & L & 1 & 20 & 3.0 $\times$ 10 $\times$  4 & 110 & 29.99808 & 30.12600 & 1.93 & 2.252 \\ 
1119 & 1342246671 & Neptune  &  -12.3 & 2012 Jun 06 & 01:55:17 &  286 & G/R & H & 1 & 20 & 3.0 $\times$ 10 $\times$  4 &  70 & 29.99739 & 29.75976 & 1.92 & 2.280 \\  
1119 & 1342246672 & Neptune  &  -12.3 & 2012 Jun 06 & 02:13:13 &  286 & G/R & H & 1 & 20 & 3.0 $\times$ 10 $\times$  4 & 110 & 29.99738 & 29.75956 & 1.92 & 2.280 \\  
1119 & 1342246673 & Neptune  &  -12.3 & 2012 Jun 06 & 02:19:03 &  286 & B/R & L & 1 & 20 & 3.0 $\times$ 10 $\times$  4 &  70 & 29.99738 & 29.75949 & 1.92 & 2.280 \\  
1119 & 1342246674 & Neptune  &  -12.3 & 2012 Jun 06 & 02:24:52 &  286 & B/R & L & 1 & 20 & 3.0 $\times$ 10 $\times$  4 & 110 & 29.99738 & 29.75943 & 1.92 & 2.280 \\  
1287 & 1342255709 & Neptune  &    1.6 & 2012 Nov 21 & 08:59:35 &  286 & G/R & H & 1 & 20 & 3.0 $\times$ 10 $\times$  4 &  70 & 29.99197 & 29.95865 & 1.91 & 2.264 \\  
1287 & 1342255710 & Neptune  &   -0.8 & 2012 Nov 21 & 09:09:23 &  286 & G/R & H & 1 & 20 & 3.0 $\times$ 10 $\times$  4 & 110 & 29.99197 & 29.95877 & 1.91 & 2.264 \\  
1287 & 1342255711 & Neptune  &   -0.8 & 2012 Nov 21 & 09:15:19 &  286 & B/R & L & 1 & 20 & 3.0 $\times$ 10 $\times$  4 &  70 & 29.99197 & 29.95884 & 1.91 & 2.264 \\  
1287 & 1342255712 & Neptune  &   -0.8 & 2012 Nov 21 & 09:21:08 &  286 & B/R & L & 1 & 20 & 3.0 $\times$ 10 $\times$  4 & 110 & 29.99197 & 29.95891 & 1.91 & 2.264 \\  
1444 & 1342270939 & Neptune  &   24.6 & 2013 Apr 26 & 17:29:50 &  286 & G/R & H & 1 & 20 & 3.0 $\times$ 10 $\times$  4 &  70 & 29.98701 & 30.45382 & 1.71 & 2.228 \\  
1444 & 1342270940 & Neptune  &   28.3 & 2013 Apr 26 & 17:39:51 &  286 & G/R & H & 1 & 20 & 3.0 $\times$ 10 $\times$  4 & 110 & 29.98701 & 30.45372 & 1.71 & 2.228 \\  
1444 & 1342270941 & Neptune  &   28.3 & 2013 Apr 26 & 17:45:40 &  286 & B/R & L & 1 & 20 & 3.0 $\times$ 10 $\times$  4 &  70 & 29.98701 & 30.45365 & 1.71 & 2.228 \\  
1444 & 1342270942 & Neptune  &   28.3 & 2013 Apr 26 & 17:51:29 &  286 & B/R & L & 1 & 20 & 3.0 $\times$ 10 $\times$  4 & 110 & 29.98701 & 30.45359 & 1.71 & 2.228 \\  
\noalign{\smallskip}
 981 & 1342238042\tablefootmark{5} & Callisto &  -2.0 & 2012 Jan 20 & 05:29:58 & 3114 & B/R & L & 1 & 10 & 2.0 $\times$ 120 $\times$ 2 &  0 & 4.98332 & 4.85010 & 11.47 & 1.370 \\ 
 981 & 1342238043\tablefootmark{5} & Callisto &  -1.9 & 2012 Jan 20 & 06:22:55 & 2274 & B/R & L & 1 & 10 & 4.0 $\times$  60 $\times$ 2 & 90 & 4.98345 & 4.85072 & 11.47 & 1.370 \\ 
\noalign{\smallskip}
 981 & 1342238042\tablefootmark{6} & Ganymede &  -2.0 & 2012 Jan 20 & 05:29:58 & 3114 & B/R & L & 1 & 10 & 2.0 $\times$ 120 $\times$ 2 &  0 & 4.97487 & 4.84122 & 11.50 & 1.499 \\ 
 981 & 1342238043\tablefootmark{6} & Ganymede &  -1.9 & 2012 Jan 20 & 06:22:55 & 2274 & B/R & L & 1 & 10 & 4.0 $\times$  60 $\times$ 2 & 90 & 4.97508 & 4.84194 & 11.50 & 1.499 \\ 
\noalign{\smallskip}
1138 & 1342247418 &  Titan   &  -18.3 & 2012 Jun 25 & 15:22:31 &  286 & B/R & H & 1 & 20 & 3.0 $\times$ 10 $\times$ 4 &  70 & 9.74085 & 9.36711 & 5.73 & 0.758 \\  
1138 & 1342247419 &  Titan   &  -18.3 & 2012 Jun 25 & 15:28:20 &  286 & B/R & H & 1 & 20 & 3.0 $\times$ 10 $\times$ 4 & 110 & 9.74087 & 9.36719 & 5.73 & 0.758 \\  
1138 & 1342247420 &  Titan   &  -18.3 & 2012 Jun 25 & 15:34:09 &  286 & G/R & H & 1 & 20 & 3.0 $\times$ 10 $\times$ 4 &  70 & 9.74088 & 9.36726 & 5.73 & 0.758 \\  
1138 & 1342247421 &  Titan   &  -18.3 & 2012 Jun 25 & 15:39:58 &  286 & G/R & H & 1 & 20 & 3.0 $\times$ 10 $\times$ 4 & 110 & 9.74089 & 9.36733 & 5.73 & 0.758 \\  
\noalign{\smallskip}\hline\noalign{\smallskip}
\multicolumn{15}{l}{{\bf Notes.}}\\
\multicolumn{15}{l}{$^{\small (1)}$\,for Uranus and Neptune taken from "diam.dat", provided by R.\ Moreno, Jul.\ 25, 2012; otherwise from JPL Horizons.}\\
\multicolumn{15}{l}{$^{\small (2)}$\,central part of the source is clearly saturated.}\\
\multicolumn{15}{l}{$^{\small (3)}$\,observations are made in fixed mode without tracking, Neptune is located fully inside the field of view, but toward one corner of the map.}\\
\multicolumn{15}{l}{$^{\small (4)}$\,data are labeled "FAILED" in the HSA.}\\
\multicolumn{15}{l}{$^{\small (5)}$\,observations are made in fixed mode without tracking, Callisto is located off center.}\\
\multicolumn{15}{l}{$^{\small (6)}$\,observations are made in fixed mode without tracking, Ganymede is located off center.}\\
\end{longtable}
\end{landscape}

\begin{sidewaystable*}
\caption[]{Additional Herschel-PACS photometer scan-map observations related to the bright planets and planetary satellites (proposals OT1\_ddan01\_1),
        taken in large-scan observing mode without SSO tracking (fixed mode), and high gain. The planets were located at the edge or outside the observed
        field of view. 
        SAA: solar aspect angle; dur.: duration of observation in seconds; fil.: filter/band combination
        (B: 70/160\,$\mu$m; G: 100/160\,$\mu$m); G.: low (L) or high (H) gain;
        R.: repetition of entire scan map; 
        S.: satellite scan speed: 20$^{\prime\prime}$/s or 60$^{\prime\prime}$/s;
        scan-map parameters: Len: scan-leg length (in arc\,min)
        $\times$ n: number of scan legs $\times$ sep: scan-leg separation (in arc\,sec);
        ang.: satellite scan angle in degrees with respect to instrument reference frame;
        r: heliocentric distance; $\Delta$: Herschel-centric distance; $\alpha$: phase angle;
        diam: effective angular diameter.}
\label{tbl:pacs_scanmap2}                                                                               
\begin{tabular}{rllrlrclrccrrrrr}
\hline\noalign{\smallskip}
   &       &        & SAA          & UTC        & dur.\ &       &      &    & S.                     & Len$\times$n$\times$sep                               & ang.        & r    & $\Delta$ & $\alpha$     & diam \\
OD & OBSID & Target & [$^{\circ}$] & Start time & [s]   & fil.\ & G. & R. & [$^{\prime \prime}$/s] & [$^{\prime}$ $\times$ \# $\times$ $^{\prime \prime}$]  & [$^{\circ}$] & [AU] & [AU]     & [$^{\circ}$] & [$^{\prime \prime}$] \\
\noalign{\smallskip}\hline\noalign{\smallskip}
 798 & 1342224620 & Saturn  &   16.5 & 2011 Jul 20 19:02:46 & 1732 & B/R & H & 4 & 20 &  15 $\times$  1 $\times$  2 &  45 & 9.64468 &  9.87529 & 5.87 & 16.829 \\  
 798 & 1342224621 & Saturn  &   16.5 & 2011 Jul 20 19:33:38 & 1732 & B/R & H & 4 & 20 &  15 $\times$  1 $\times$  2 & 135 & 9.64469 &  9.87569 & 5.86 & 16.829 \\  
 798 & 1342224622 & Saturn  &   16.5 & 2011 Jul 20 20:08:29 & 2325 & B/R & H & 1 & 20 &  40 $\times$  1 $\times$  2 &  45 & 9.64469 &  9.87611 & 5.86 & 16.828 \\  
 798 & 1342224623 & Saturn  &   16.9 & 2011 Jul 20 20:48:17 & 2325 & B/R & H & 1 & 20 &  40 $\times$  1 $\times$  2 & 135 & 9.64470 &  9.87656 & 5.86 & 16.827 \\  
 807 & 1342224837 & Saturn  &   24.3 & 2011 Jul 29 13:14:30 & 1732 & B/R & H & 4 & 20 &  15 $\times$  1 $\times$  2 &  45 & 9.64724 & 10.01382 & 5.58 & 16.597 \\  
 807 & 1342224838 & Saturn  &   24.3 & 2011 Jul 29 13:45:40 & 1732 & B/R & H & 4 & 20 &  15 $\times$  1 $\times$  2 & 135 & 9.64725 & 10.01414 & 5.57 & 16.596 \\  
 807 & 1342224839 & Saturn  &   24.3 & 2011 Jul 29 14:20:31 & 2325 & B/R & H & 1 & 20 &  40 $\times$  1 $\times$  2 &  45 & 9.64726 & 10.01457 & 5.57 & 16.595 \\  
 807 & 1342224840 & Saturn  &   24.7 & 2011 Jul 29 15:00:19 & 2325 & B/R & H & 1 & 20 &  40 $\times$  1 $\times$  2 & 135 & 9.64726 & 10.01500 & 5.57 & 16.595 \\  
\noalign{\smallskip}
 969 & 1342236891 & Uranus  &   15.7 & 2012 Jan 07 11:26:54 & 3061 & B/R & H & 3 & 20 &  25 $\times$  1 $\times$  2 &  45 & 20.07598 & 20.31566 & 2.73 & 3.469 \\ 
 969 & 1342236892 & Uranus  &   15.8 & 2012 Jan 07 12:20:10 & 3061 & B/R & H & 3 & 20 &  25 $\times$  1 $\times$  2 & 135 & 20.07598 & 20.31629 & 2.73 & 3.469 \\ 
 975 & 1342237436 & Uranus  &   21.8 & 2012 Jan 13 11:59:55 & 3061 & B/R & H & 3 & 20 &  25 $\times$  1 $\times$  2 &  45 & 20.07571 & 20.41527 & 2.64 & 3.452 \\ 
 975 & 1342237437 & Uranus  &   21.7 & 2012 Jan 13 12:52:30 & 3061 & B/R & H & 3 & 20 &  25 $\times$  1 $\times$  2 & 135 & 20.07571 & 20.41587 & 2.64 & 3.452 \\ 
\noalign{\smallskip}\hline
\end{tabular}
\end{sidewaystable*}

\end{appendix}

\end{document}